# Plasmon-induced strong interaction between chiral molecules and orbital angular momentum of light

Tong Wu, Rongyao Wang and Xiangdong Zhang*

School of Physics and Beijing Key Laboratory of Nanophotonics & Ultrafine Optoelectronic Systems, Beijing Institute of Technology, 100081, Beijing, China

*Correspondence to zhangxd@bit.edu.cn

ABSTRACT: Whether or not chiral interaction exists between the optical orbital angular momentum (OAM) and a chiral molecule remains unanswered. So far, such an interaction has not been observed experimentally. Here we present a T-matrix method to study the interaction between optical OAM and the chiral molecule in a cluster of nanoparticles. We find that strong interaction between the chiral molecule and OAM can be induced by the excitation of plasmon resonances. An experimental scheme to observe such an interaction has been proposed. Furthermore, we have found that the signal of the OAM dichroism can be either positive or negative, depending on the spatial positions of nanocomposites in the cross-sections of OAM beams. The cancellation between positive and negative signals in the spatial average can explain why the interaction has not been observed in former experiments.

Many biomolecules are chiral in nature. Chirality plays a pivotal role in biochemistry and the evolution of life itself[1, 2]. When chiral molecules interact with circularly polarized photons, they exhibit optical activity, such as the circular dichroism (CD) effect describing the difference in molecular absorption of right- and left-handed circularly polarized photons. The related CD spectroscopic technique has been one of the central methods for optically probing the molecular chirality[3, 4]. In addition to polarized light, which is intrinsically linked to the spin angular momentum of photons, photons can carry orbital angular momentum (OAM), which is associated with the helicity in spatial phase distribution[5-10]. Over the past decades, many researchers have studied the interactions between OAM photons and chiral molecules[11-18]. Although many of them have predicted an exchange can occur between the OAM of the photon and the center of mass motion of the molecule, whether OAM photons can interact with molecular chirality remains an open question.

Vortex beams with various OAMs have been experimentally realized in the optical domain[5, 6]. The possibility of encoding large amounts of information in vortex beams due to the absence of an upper limit has raised the prospects of their applicability in information processing tasks[7, 8]. The OAM beams have been widely used in manipulating and trapping microscopic particles[9, 10].

The problem is whether vortex beams with OAMs can be used to probe molecular chirality? Does this additional degree of freedom play a role in CD? If the interaction between the OAM and the chiral molecule was found, it is potentially useful for a broad range of research areas and applications. Thus, the electromagnetic (EM) interactions of OAM photons with atoms and molecules have received considerable attention[11-18]. Some theoretical investigations predicted that the interaction of OAM light with atoms and molecules should be observable within the electric dipole approximation[11-13], but the others obtained a contradictory result[14–16]. The latter outcome has been supported by recent experimental studies[17, 18], which verified that the influence of the OAM on CD of chiral molecules can not be detectable.

On the other hand, recent investigations have demonstrated that light-molecule interactions can be enhanced via resonant excitations of surface plasmon resonance (SPR) of metallic nanostructures. For CD probe of molecular chirality, it has been reported a large difference in absorption of right- and left-handed circularly polarized photons in the SPRs [19-27]. In addition, the SPRs have been also used to improve nanoparticle(NP)-assisted biosensing [28-30] and Raman scattering [31-33]. In this work, we attempt to explore the role of SPRs in the promotion of the interactions between the OAMs and the chiral molecules.

To address this issue, we study the interaction between the OAM beams and the nanocomposite comprising chiral molecule and metallic NPs. A method to study such an interaction has been developed by using the T-matrix. Based on such a method, we study the interaction between OAM beams and chiral molecules located in the vicinity of NPs. We find that the strong interaction between the chiral molecule and OAM can be induced by the excitation of plasmon resonances of NPs. Experimentally, such an interaction could be observed if we define the OAM dichroism (OD) to be the difference of absorption rates between two focused linear polarized OAM beams with opposite topological indices. We should stress here that a nonzero spatial-average OD signal exists only for the nanocomposites located in a certain spatial region of OAM beams. In addition, we also find that the OD can be further enhanced when the orientations of nanocomposites are fixed.

**RESULTS AND DISCUSSIONS**

We consider here a semi-classical hybrid system consisting of a molecules and a cluster of $n$ NPs, which is excited by an OAM beam with $\boldsymbol{E}_{ext} = \text{Re}\, \boldsymbol{E}_{inc\pm}^{(\pm l)}(\boldsymbol{r})e^{-i\omega t}$ and $\boldsymbol{B}_{ext} = \text{Re}\, \boldsymbol{B}_{inc\pm}^{(\pm l)}(\boldsymbol{r})e^{-i\omega t}$, $\boldsymbol{E}_{inc\pm}^{(\pm l)}$ and $\boldsymbol{B}_{inc\pm}^{(\pm l)}$ represent electronic and magnetic fields of the OAM beam (their expressions in aplanatic system are given in supplementary materials). The intensity of the incident wave is considered to be weak enough that the mechanical interaction between the light and the hybrid system can be neglected. The molecules used here are assumed to be point-like two-level systems, no vibrational structure of transitions is considered. According to the previous investigations, the master equation for quantum states of the molecule can be written as [19, 23, 25]

$$\hbar \frac{\partial \rho_{ij}}{\partial t} = i\left[\hat{\rho}, \hat{H}\right]_{ij} - \Gamma_{ij}(\rho) \quad , \tag{1}$$

where $\hat{H} = \hat{H}_0 + \hat{H}_1$ is the Hamiltonian of the molecule and $\hat{H}_0$ describes the internal electronic structure of the molecule, $\hat{\rho}$ is the density matrix and $\rho_{ij} = \sigma_{ij}e^{-i\omega t}$ is the corresponding matrix element. Here $\hat{H}_1 = -\hat{\boldsymbol{\mu}} \cdot \boldsymbol{E}_T - \hat{\boldsymbol{m}} \cdot \boldsymbol{B}_T + V_{quad}$ is the light-matter interaction operator, $\hat{\boldsymbol{\mu}} = \sum_t e_t \hat{\boldsymbol{r}}_t$ and $\hat{\boldsymbol{m}} = \sum_t \frac{e_t}{2m_t} \hat{\boldsymbol{r}}_t \times \hat{\boldsymbol{p}}_t$ are electric and magnetic operators, and $(\boldsymbol{E}_T, \boldsymbol{B}_T) = \text{Re}\,(\boldsymbol{E}_{tot}, \boldsymbol{B}_{tot})e^{-i\omega t}$ represents the EM field acting on the molecule. Where $e_t$, $m_t$, $\hat{\boldsymbol{r}}_t$ and $\hat{\boldsymbol{p}}_t$ are the electric charge, mass, position and momentum operators for the $t^{th}$ charged particle in the molecule system. The electric quadrupole interaction term $V_{quad}$ bears the same order of magnitude with the magnetic dipole term $-\hat{\boldsymbol{m}} \cdot \boldsymbol{B}_T$, whose expression is written as $V_{quad} = -\frac{1}{3}\sum_{\alpha,\beta} Q_{\alpha,\beta} \frac{\partial E_{T\alpha}}{\partial r_\beta}$, with $Q_{\alpha,\beta}$ being the element of the electric quadrupole operator. In this work we set the electric dipole of the molecule to be aligned with the rotational symmetry axis of the nano system. Thus, this term makes no contribution to the dichroism and can be omitted here. $\Gamma_{ij}(\rho)$ is the relaxation term which describes the damping with $\Gamma_{11}(\rho) = -\Gamma_{22}(\rho) = -\gamma_{22}\rho_{22}$, $\Gamma_{12}(\rho) = \gamma_{21}\rho_{12}$ and $\Gamma_{21}(\rho) = \gamma_{21}\rho_{21}$. By solving this equation within rotating-wave approximation, and letting $\rho_{21} = \sigma_{21}e^{-i\omega t}$, we can obtain energy absorption rate of the system:

$$Q = Q_{mol} + Q_{NP} \qquad (2)$$

where $Q_{mol}$ represents the absorption rate of the molecule in the system, which is expressed as

$$Q_{mol} = \frac{\omega_0 \gamma_{21}}{2} \frac{|\mu_{21} \cdot E' + m_{21} \cdot B'|^2}{|\hbar(\omega-\omega_0) + i\gamma_{21} - G|^2}, \qquad (3)$$

where $\omega_0$ is the frequency of molecular transition, $\gamma_{21}$ is the relaxation term, $\mu_{21} = \langle 2|\hat{\mu}|1\rangle$ and $m_{21} = \langle 2|\hat{m}|1\rangle$ are the matrix elements of electric and magnetic dipole momentum operators. $(E', B') = (E_{inc\pm}^{(\pm l)} + E_s, B_{inc\pm}^{(\pm l)} + B_s)$, with $E_s$ and $B_s$ being the EM field of scattered wave by the NPs in the absence of the molecule. $G$ is the function which describes the broadening of the resonance peak of the signal due to the interactions between the molecules and NPs, and its expression can be founded in Ref. 25. The induced dipole of the molecule is given by $d = 2\text{Re}\{\sigma_{21}\mu_{12}e^{-i\omega t}\}$. $Q_{NP}$ in Eq. (2) is the absorption rate of NPs in the system. According to the law of energy conservation, it equals to the energy flows into them. Thus we have

$$Q_{NP} = -\frac{1}{2}\text{Re}\sum_{\xi=1}^{NP}\iint_{\Omega_\xi} \mathbf{n}\cdot \mathbf{E}_{tot}(r-r_\xi)\times \mathbf{H}^*_{tot}(r-r_\xi)d\Omega_\xi, \qquad (4)$$

where $(\boldsymbol{E}_{tot}(r), \boldsymbol{H}_{tot}(r))$ are the amplitudes of the total complex EM field in the space, and $\Omega_\xi$ denotes a surface circumscribing around the $\xi th$ NP. If we define OAM dichroism (OD) of the system as the difference of absorption rate between two focused linear polarized OAM beams with opposite topological indices $+l$ and $-l$, at the same time, they are space inversion of each other, it can be expressed as

$$OD = OD_{mol} + OD_{NP}, \tag{5}$$

where $OD_{mol} = Q_{mol}(+l) - Q_{mol}(-l)$ represents the contribution of the molecule in the molecule-NP nanocomposites, whose value is influenced by the NPs. $OD_{NP} = Q_{NP}(+l) - Q_{NP}(-l)$ is the molecule-induced OD for the nanoparticles, where $Q_{mol}(\pm l)$ and $Q_{NP}(\pm l)$ can be written as (see methods section for detail)

$$Q_{NP/mol}(\pm l) = \operatorname{Re} \boldsymbol{a}(\pm l) R^\dagger(\alpha,\beta,\gamma) \bar{\bar{M}}_{NP/mol} R(\alpha,\beta,\gamma) \boldsymbol{a}^\dagger(\pm l), \tag{6}$$

where $\boldsymbol{a}(\pm l)$ is arranged as $\boldsymbol{a}(\pm l) = \{ a_{1\pm}^{\pm l} \; b_{1\pm}^{\pm l} \; \cdots \; a_{\nu\pm}^{\pm l} \; b_{\nu\pm}^{\pm l} \; \cdots \}$, with $a_{\nu\pm}^{\pm l}$ and $b_{\nu\pm}^{\pm l}$ are set to be the vector spherical function (VSF) expansion coefficients of an OAM beam propagating along the $z$ direction (see supplementary material). $\nu = (m,n)$ is introduced here for notation simplification, and the convention $\nu = 1,2,....,$ when $n = 1,2,...,$ and $m = -n,...,n$ is used. Absorption of waves in any direction can be calculated by changing the Euler angles $\alpha$, $\beta$ and $\gamma$. $R(\alpha,\beta,\gamma)$ is the rotation matrix which depends only on the direction of the incident wave which is given in Refs. 34-36, while $\bar{\bar{M}}_{NP/mol}$ is a matrix determined only by the property of the illuminated system, which has the form of

$$\bar{\bar{M}}_{NP/mol} = \begin{bmatrix} & \vdots & & \\ \cdots & [M_{NP/mol}]_{\nu\nu'}^{11} & [M_{NP/mol}]_{\nu\nu'}^{12} & \cdots \\ & [M_{NP/mol}]_{\nu\nu'}^{21} & [M_{NP/mol}]_{\nu\nu'}^{22} & \\ & \vdots & & \end{bmatrix}. \tag{7}$$

Based on Eqs. (4-7), we can obtain OD through numerical calculations. However, the results depend on geometrical configuration of the system. For example, the OD signals for the NP-molecule system being fixed in a certain region in space but possessing random orientations are different from those for the case being distributed uniformly in the whole space with random orientations. They are also different from the case, which the direction of the chiral system is fixed, and the system is distributed uniformly in some certain regions. In the following, we will deal with each case separately.

**OD for nanocomposites with fixed positions and random orientations.** In practice, molecule-NP nanocomposites in liquid environment have random orientations, averaging over the solid angles of the directions of the samples need to be done. This is equivalent with averaging over the solid angles of the directions of the incident light as has been considered for the calculation of absorption and CD spectra [19, 23, 25-26]. Likewise, we did the average OD calculations for a system with random orientations but fixed position. The absorption for the OAM light by nanoparticles and the molecule should be rewritten as

$$\langle Q_{NP/mol}(\pm l) \rangle = \boldsymbol{a}(\pm l) \langle M_{NP/mol} \rangle_\Omega \boldsymbol{a}^\dagger(\pm l) \tag{8}$$

with

$$\left[ \langle M_{NP/mol} \rangle_\Omega \right]^{ij}_{mn,m_1n_1} = \frac{1}{2n+1} \sum_{m'=-n}^{n} [M_{NP/mol}]^{ij}_{m'n,m'n_1} \delta_{mm_1} \delta_{nn_1} \;, \tag{9}$$

where $\langle \cdots \rangle_\Omega$ represents the averaging over the solid angle. For a chiral system, $\left[ \langle M_{NP/mol} \rangle_\Omega \right]^{12}_{vv}$ or $\left[ \langle M_{NP/mol} \rangle_\Omega \right]^{21}_{vv}$ should not be vanished. Also by analyzing the properties of $a^{\pm l}_{v+}$ and $b^{\pm l}_{v+}$ under space inversion, the orientation averaged OD of the system is given by (see methods section for detail)

$$\langle OD_{NP/mol} \rangle_\Omega = \text{Re}\left[ a \langle M_{NP/mol} \rangle_\Omega a^\dagger - \tilde{a} \langle M_{NP/mol} \rangle_\Omega \tilde{a}^\dagger \right]$$

$$= 2\sum_{v=1}^{\infty} \text{Re}\left\{ \left[\langle M_{NP/mol} \rangle_\Omega\right]_{vv}^{12} a_{v+}^l b_{v+}^{l\,*} + \left[\langle M_{NP/mol} \rangle_\Omega\right]_{vv}^{21} b_{v+}^l a_{v+}^{l\,*} \right\}, \quad (10)$$

$$\langle OD \rangle_\Omega = \langle OD_{mol} \rangle_\Omega + \langle OD_{NP} \rangle_\Omega$$

$$= 2\sum_{v=1}^{\infty} \text{Re}\left\{ \left[\langle M \rangle_\Omega\right]_{vv}^{12} a_{v+}^l b_{v+}^{l\,*} + \left[\langle M \rangle_\Omega\right]_{vv}^{21} b_{v+}^l a_{v+}^{l\,*} \right\} \quad (11)$$

with $\langle M \rangle_\Omega = \langle M_{NP} \rangle_\Omega + \langle M_{mol} \rangle_\Omega$. From the above equations, the orientation averaged OD of the system can be obtained.

Figure 1(b) and (c) show the calculated orientation averaged OD as a function of wavelength for a chiral molecule being placed in the vicinity of a single gold sphere (Fig.1(a)) for the OAM incident beam with $l=\pm 1$ and $l=\pm 2$, respectively. Here the OAM beams used are x directional polarized Laguerre-Gaussian (LG) lights with a numerical aperture of $NA = \sqrt{\varepsilon_r} \sin\alpha_{max} = 0.55$ and having a filling factor of $f_0 = \omega_0/(f \sin\alpha_{max}) = 1$. The focusing position of the beam locates at the origin of the coordinate, as shown in Fig. 1(a). The radius of the gold sphere is taken as 15nm, the distance between the chiral molecular and the sphere is taken as 2nm. The parameters of the molecular dipole are taken according to Refs. 1 and 23: $\mu_{12} = |e|r_{12}$ and $\vec{\mu}_{12} \cdot \vec{m}_{21}/\mu_{12} = i|e|r_0\omega_0 r_{21}/2$ with a resonance wavelength of 300nm. The direction of the electric dipole of the molecule is set to be align with the symmetry axis of the NP. In our calculations, we use $r_{12} = 2\text{Å}$, $r_0 = 0.05\text{Å}$ and $\gamma_{12} = 0.3eV$. For the dielectric functions of Au, the Johnson's data were adopted [37], the permittivity of water is taken to be $\varepsilon_0 = 1.8$. If the OD signals are in units of $M^{-1}cm^{-1}$, they have to be multiplied by $NA/(0.23I)$ where $NA = 6.02\times 10^{23}$ is Avogadro's number [19, 23], $I$ is related to the square of the incident electric field through $I = \varepsilon_{vac} c\sqrt{\varepsilon_r} \left|E_{inc+}^{(+l)}\right|^2/2$.

The green line and red line in Fig. 1 correspond to $OD_{mol}$ and $OD_{NP}$, respectively, the total OD is described by the blue line. The extinction cross section of the gold sphere is plotted as the orange line, and the plasmon resonance peak appears at the wavelength of 520nm. Comparing the present OD spectra with that of pure chiral molecular (without the NP, insets of Fig. 1(b) and (c)), we find that the $OD_{mol}$ signal around $\lambda = 300nm$ is improved about 2 times. Moreover, a new OD band appears in the plasmon resonance spectral region. These results reveal not only a plasmon-enhanced chiral molecule-OAM interaction at the molecule resonance frequency, but also an OD effect at the plasmonic resonant frequency of metallic NP. Note that spherical NP itself is achiral, such an induced plasmonic OD correlated with the enhanced interaction between the chiral molecular and the OAM has not been reported in previous studies.

For a single sphere, the enhancement of OD signal by the NP is small and weak. When the chiral molecule is put in the hotspot of the dimer, as shown in Fig. 1(d), giant enhancement effect can be observed. The cases when the molecule is put in the gap of an Au dimer with an inter particle distance of 1nm is presented in Fig. 1(e) and (f) for the OAM beam with $l = \pm 1$ and $l = \pm 2$, respectively. Comparing it with the case of the single sphere, the total $OD$ is significantly enhanced near the wavelength of coupled plasmon resonance. Specifically, the plasmonic $OD$ at the hotspot can be 50 times larger than that of the single sphere when $l = \pm 1$, while the enhancement of $OD_{mol}$ can also reach 30 times. The phenomena are similar for the OAM beam with different topological indices such as $l = \pm 2$ illustrated in Fig. 1(f).

Considering that the EM field of the OAM beam is nonuniform in space, we investigate the OD signal at different position of the OAM beams. Fig. 2(a) and (b) correspond to the case of single sphere with $l = \pm 1$ and $l = \pm 2$, respectively, Fig. 2(c) and (d) to the case of dimer. The olive/red/black blue line in Fig. 2 represents the orientation averaged OD when the molecule is put at $x$ =50nm/100nm/150nm in the focal plane (origin of the coordinate is set to be the focal center of the OAM beams). The corresponding OD for the mirror reflected samples are also plotted as green, pink and sky blue lines, as expected these signals are of same values but opposite signs. With the position of the nanocomposite is shifted away from the beam center, the OD signal decreases. In order to disclose such a phenomenon we rewrite Eq.(10) under the dipole approximation, which means the cut-off of $v$ is set to be 1. If we consider the nanocomposite system being located not too close to the beam center, the electric and magnetic fields illuminated on it can be viewed as a constant vector. Using Eqs.(8) and (11), and expand the incident wave according to Eqs. (42-43), the OD can be rewritten as

$$\text{OD} \approx C \left\{ -\text{Re}\left[ \left[\langle M \rangle_\Omega \right]^{12}_{(01,01)} + \left[\langle M \rangle_\Omega \right]^{21}_{(01,01)} \right] \text{Im}\left( \boldsymbol{E}^{(+l)}_{inc+} \cdot \boldsymbol{H}^{(+l)*}_{inc+} \right) \right.$$
$$\left. - \text{Im}\left[ \left[\langle M \rangle_\Omega \right]^{12}_{(01,01)} - \left[\langle M \rangle_\Omega \right]^{21}_{(01,01)} \right] \text{Im}\left( -i\boldsymbol{E}^{(+l)*}_{inc+} \cdot \boldsymbol{H}^{(+l)}_{inc+} \right) \right\} , \quad (12)$$

where $C$ is a real constant. It is worth to note when the molecule is put inside the gap of the dimer, Eq.(12) is still valid if $\left[\langle M \rangle_\Omega \right]^{12}_{(01,01)}$ and $\left[\langle M \rangle_\Omega \right]^{21}_{(01,01)}$ are solved beyond the dipole approximation (see methods section for detail). Our calculations show that the first term in Eq.(12) plays a leading role in the OD signals, thus the signal is proportional to $\text{Im}[\boldsymbol{E}^{(+l)}_{inc+} \cdot \boldsymbol{H}^{(+l)*}_{inc+}]$. This variable has been firstly introduced by Tang and Cohen to characterize the chirality of light[38], since they are proportional to the difference of absorption rates between two oppositely handed molecules. Later it was Cameron et al.[39] who indicated that this term is in fact proportional to the helicity density which describes the 'screw action' of the EM field[40]. The calculated results for $\text{Im}[\boldsymbol{E}_{inc+}^{(+l)} \cdot \boldsymbol{H}_{inc+}^{(+l)*}]/|\boldsymbol{E}_{inc+}^{(+l)}|^2$ as a function of position are plotted in Fig. 2(e) and (f) for the OAM beam with $l=1$ and $l=2$ at the wavelength of 300nm, respectively. It can be found the value of $\text{Im}[\boldsymbol{E}_{inc+}^{(+l)} \cdot \boldsymbol{H}_{inc+}^{(+l)*}]/|\boldsymbol{E}_{inc+}^{(+l)}|^2$ is close to that of the circular polarized wave, of which value decreases with the increase of the relative distance between nanocomposites and the center of the beam, which directly leads to the decrease of the OD signal. This phenomenon has also been reported in Ref. 41, where a large helicity density has been observed at the center of the Bessel beam with the OAM.

We would like to point out that the above OD signals originate from the interaction of the OAM and the chiral molecule, which are not because of the spin angular momentum, because we have used the linear polarized OAM light as the incident wave with no spin angular moment even when it is strongly focused by lens [42, 43].

The previous theoretical investigations have shown that only the spin angular momentum leads to a differential absorption, there is no interaction of the OAM of the beam with the chirality of the molecule [14-15,44-45]. This is because the signals calculated in these works are the average of contributions from all the molecules in spaces. In fact, our calculated results have also shown that the OD signals do not exist in the case of the whole spatial average either, which will be discussed in the following section.

**OD for nanocomposites with whole spatial average.** Because of heterogeneity of the OAM beam in space, the spatial average for the OAM dichroism is needed in some cases, for example, to compare the theoretical results with the experimental measurements as described in Refs. 18, 19. Here we consider a spatial averaged OD performed in the focal plane normal to the propagation direction of the OAM beam, which can be expressed as

$$\langle \mathrm{OD}(z_0)\rangle_\Omega^S = \int_{\rho_0=0}^{\infty}\int_{\phi=0}^{2\pi}\langle \mathrm{OD}(\rho_0,z_0)\rangle_\Omega \rho_0 d\rho_0 d\phi \ , \tag{13}$$

where $\langle \mathrm{OD}(\rho_0,z_0)\rangle_\Omega$ is the angular averaged OD signal of the system which is located at $\rho_0 = (x_0, y_0, z_0)$. Substituting Eq.(10) into Eq.(13), we find

$$\langle \mathrm{OD}(z_0)\rangle_\Omega^S = 2\mathrm{Re}\left\{ \left[\langle M\rangle_\Omega\right]_{vv}^{12} \int_{\rho_0=0}^{\infty}\int_{\phi=0}^{2\pi} a^l_{v+} b^{l\,*}_{v+} \rho_0 d\rho_0 d\phi + \left[\langle M\rangle_\Omega\right]_{vv}^{21} \int_{\rho_0=0}^{\infty}\int_{\phi=0}^{2\pi} b^l_{v+} a^{l\,*}_{v+} \rho_0 d\rho_0 d\phi \right\}. \tag{14}$$

The involved integration in Eq.(14) can be calculated analytically by using the integrations given in (supplementary materials). If we use $s$ and $s'$ to stand for $a$ and $b$, the involved integration in Eq.(14) can be expressed as

$$\int_{\rho_0=0}^{\infty}\int_{\phi=0}^{2\pi} s_v^* s'_v \rho_0 d\rho_0 d\phi = \int_0^\alpha \int_0^{\alpha'} C_{ss'}^{(v)}(\alpha,\alpha',p,|l|) \sum_{i,j=1}^{2} \left[ A^*_{s\,i+}A_{s'\,j+} + A^*_{s\,i-}A_{s'\,j-} \right]$$
$$\left[ I_1^{(ml)}\Theta_i(\alpha)\Theta_j(\alpha') + I_2^{(ml)}\Theta_i(\alpha)\Omega_j(\alpha') \right. \tag{15}$$
$$\left. + I_3^{(ml)}\Omega_i(\alpha)\Theta_j(\alpha') + I_4^{(ml)}\Omega_i(\alpha)\Omega_j(\alpha') \right]d\alpha d\alpha'$$

with

$$C_{ss'}^{(v)}(\alpha,\alpha',p,|l|) = 8\pi\eta_f^2 \left[\gamma_{mn}\frac{n_1}{n_2}\right]\sqrt{\cos\alpha\cos\alpha'}\sin\alpha\sin\alpha'$$
$$e^{ikz_0(\cos\alpha-\cos\alpha')}\mathsf{P}_{pl}(\sin\alpha')\mathsf{P}_{pl}(\sin\alpha)e^{-f^2(\sin^2\alpha+\sin^2\alpha')/\omega_0^2} \tag{16}$$

where $\Theta_1 = \Omega_2 = \pi_{mn}$ and $\Theta_2 = \Omega_1 = \tau_{mn}$. The terms $A_{s\,i+}$, $A_{s'\,j+}$, $A_{s\,i-}$ and $A_{s'\,j-}$ are polarization dependent coefficients, their expressions are written as

$$A_{a1\pm} = A_{b2\pm} = \pm 0.5 p_x - 0.5 i p_y$$
$$A_{a2\pm} = A_{b1\pm} = 0.5 p_x \mp 0.5 i p_y \tag{17}$$

The expressions for $I_1^{(ml)}, I_2^{(ml)}, I_3^{(ml)} = -I_2^{(ml)}, I_4^{(ml)}$ and the meaning for all the involved variables have been given in supplementary materials. Substitute Eq.(14) into Eq.(12), using the condition $I_3^{(ml)} = -I_2^{(ml)}$, it can be founded that $\int_{\rho_0=0}^{\infty}\int_{\phi=0}^{2\pi} a^{l*}_{v+} b^l_{v+} \rho_0 d\rho_0 d\phi = \int_{\rho_0=0}^{\infty}\int_{\phi=0}^{2\pi} b^{l*}_{v+} a^l_{v+} \rho_0 d\rho_0 d\phi = 0$ for linear polarized beams

($\text{Imag}[p_x/p_y]=0$). Thus from Eq.(14), we can see that $\langle OD(z_0)\rangle^S_\Omega \equiv 0$ for any linear polarized OAM beam.

In order to analyze the physical origin of the phenomenon, we plotted in Fig. 3(a) the $\langle OD(\rho_0,z_0)\rangle_\Omega$ distribution in the focal plane at the wavelength of 300nm for the molecule-NP system described in Fig. 1(a). The X-polarized OAM beam with $NA=0.55$ and $l=+1$ is used. The corresponding intensity distribution of the incident electric field is plotted in Fig. 3(b) for comparison. One can see that the OD signal of this molecule-NP system change sign when it is moved away from the beam center. The cancellation between positive and negative signals in the spatial average explains why $\langle OD(z_0)\rangle^S_\Omega$ is zero. Here it is worth to note that even though the electric field intensity in the central region is weak, the OD signal in this region is still large due to the higher helicity density near the beam center.

In fact, from the above theory, the same conclusion can be drawn for the single molecule without NPs. This may explain why the influence of the OAM on the CD of chiral molecules has not been observed in the previous experiments [17, 18]. Our theory has demonstrated that the *OD* signal from the chiral molecule can not be observed in the whole spatial average even with the aid of plasmonic NPs. It is worth to note that $\langle OD(z_0)\rangle^S_\Omega \equiv 0$ is also true for systems of pure NPs since Eq.(10) is valid in the absence of the molecule. Thus such an OD signal can not be observed for the structural chirality [25, 46-53]. Although the OD signal does not exist due to the whole spatial average, we can still observe the signal at some conditions. In the following, we will discuss such phenomena.

**OD for nanocomposites with fixed orientations in a defined region.** We consider some NPs locate in some defined regions, as shown in Fig. 4(a). The samples are concentrated in some circular areas. In these areas, the samples have fixed axis relative to the incident waves. In a real experiment, this corresponds to the situation where the molecules or the composite systems are fixed on the substrate. Here the direction of the molecule's electric dipole is set to be aligning with the incident OAM wave and the symmetry axis of the composite system. If the incident wave is oblique, the OD signal appears even when the structure is achiral. This effect also occurs for circular polarized waves known as extrinsic dichroism caused by mutual orientations of incident waves and samples[54-56]. In the following, we consider the cases where the OD signals are gathered from circular areas around the beam center under the vertical incidence (see Fig 4(b)). Specifically the OD signal is rewritten as

$$\Delta(\rho_s, \rho_e) = \int_{\rho=\rho_s}^{\rho_e} \int_{\phi=0}^{2\pi} \left[ Q_{+l}(\rho,\phi,z) - Q_{-l}(\rho,\phi,z) \right] \rho d\rho d\phi \quad , \tag{18}$$

When setting $\rho_s = 0$ and $\rho_e = \infty$, Eq.(18) also gives the result for the whole spatial average, it is $\Delta \equiv 0$ for all the cases (single molecule or molecule-NP systems), which is the same with the orientation averaged case discussed above. However, if the spatial average is calculated for some defined regions, the situation becomes different. The olive, red and dark blue lines in Fig. 4 are the calculated results for the orientation fixed OD that are averaged in some spatial regions as shown in Fig. 4(b), which correspond to the region $\rho = 0 - 50nm$, $50 - 100nm$ and $100 - 150nm$, respectively. The corresponding OD for the samples with opposite chirality are shown by green, pink and sky blue lines. The results are calculated from Eq.(18) numerically, and multiplied by a factor of $NA/(0.23I)$ for normalization. Here $I$ is equal to $\int_{\rho_0=\rho_s}^{\rho_e} \int_{\phi=0}^{2\pi} \varepsilon_{vac} c \sqrt{\varepsilon_r} \left| E_{inc+}^{(+l)} \right|^2 / 2 \rho_0 d\rho_0 d\phi$.

Figure 4(c) and (d) show the calculated results for the nanocomposite consisting of an Au nanoparticle and a chiral molecule under the OAM incident beam with $l = \pm 1$ and $l = \pm 2$, respectively, while (e) and (f) are for the system with Au dimer and a chiral molecule. The parameters of nanocomposites are taken identical with those in Fig..1. The corresponding OD signals for the single molecule without NPs are plotted in the insets of Fig..4(c) and (d). In contrast to the case under the whole spatial average, we find that strong plasmon-induced OD signals appear for the region spatial average. Generally they decrease with the increase of the relative distances between nanocomposites and the center of the beam, that is, the maximum appears in the region $\rho = 0-50 nm$. However, it is different for the case in Fig. 4(f) for the dimer system with $l=\pm 2$, the maximum of OD signals appear in the region $\rho = 50-100 nm$. This is because the relative value of the longitude part of the electric field $\left| E_{inc+z}^{(+l)}(\rho,\phi) / E_{inc+}^{(+l)}(\rho,\phi) \right|$ at various wavelengths has different behaviors for the OAM incident beam with $l=\pm 1$ and $l=\pm 2$.

In Fig. 4(g) and (h), we plot the value of $\int_0^{2\pi} \left| E_{inc+z}^{(+l)}(\rho,\phi) / E_{inc+}^{(+l)}(\rho,\phi) \right| d\phi / (2\pi)$ at the focal plane as a function of the wavelength and the relative distances between the center of the x-polarized OAM beam and nanocomposites for the case with $l=1$ and $l=2$, respectively. The relative values of the longitude electric field in the region $\rho = 0-50 nm$ is obviously larger than those in other regions for the case with $l=\pm 1$. This is in contrast to the case with $l=\pm 2$, where the maximum of the longitude electric field appears in the region $\rho = 50-100 nm$. The magnitude of the longitude electric field determines the field intensity in the hotspots of the dimer. In panels A and B of Fig.4, we plot the intensity distribution of the electric field in the dimer under the OAM incident beam with $l=2$ at the wavelength of 605nm. Panel A corresponds to the case where the dimer is put at $x=100 nm$, while Panel B is for the same system situated at the center of the beam. Since the OAM beam with $l=2$

has a vanishing $E_z$ at the center of the beam [57], no hotspot is generated. This means that the electric field is not greatly amplified in the position of the molecule.

The calculated results shown in Fig. 4 are only for the case under the spatial average in some defined regions when the orientations of nanocomposites are fixed. In fact, if we consider random orientations of nanocomposites and do orientation average for such a case again, the OD signals still exist. Fig. 5 displays the calculated results. The results in Fig. 5(a), (b), (c) and (d) correspond to those in Fig. 4(c), (d), (e) and (f), respectively. Comparing them, we find that the OD signals decrease when the orientation averages are performed, however, plasmon-induced OD signals are still large.

**CONCLUSIONS**

In summary, we have developed a T-matrix method to study the interaction between optical OAM and the chiral molecule in a cluster of nanoparticles. Our results have revealed that the strong interaction between the chiral molecule and OAM can be induced by the excitation of plasmon resonances. Such an interaction leads to the OAM dichroism effect, which depends on the geometrical configuration of the molecule-NP system, such as the orientations and spatial positions of nanocomposites, in the illumination of OAM beam. It is important to note that the sign of the OAM dichroism signal can be either positive or negative, depending on the spatial positions of the nanocomposite in the cross-section of OAM beams. In this point, experimental observation of a nonzero OAM dichroism signal from the molecule-NP system is challenging, since it requires spatial arrangement of nanocomposites in a very limited space region of the incident OAM beam. This leads to the OAM wave improper for the detection of the chirality of samples which are randomly distributed. However, it does not prevent the OAM beam from becoming as an alternative probe of the chirality of an individual nano structure. Since the OAM adds another dimension for the judgment of the chirality of nano structure in addition to the spin angular momentum, we believe it may play an important role in the realization and improvement of chirality detection at the nanoscale[55,58]. Our theoretical study reveals here the possibility to observe the interactions between chiral molecules and OAM beams, which paves the way to explore a novel plasmon-based spectroscopic technique for optically detecting the molecular chirality.

**METHODS**

**T-matrix formula for the absorption rate.** In this section, we provide the T-matrix formula for the absorption rate. The incident wave can be expanded as a series of VSFs in a coordinate system of which origin is set to the position of the molecule:

$$E_0(r) = \sum_{v,v'=0}^{\infty} [a_v \ b_v] R_{vv'}(-\gamma,-\beta,-\alpha) \begin{bmatrix} M_{v'}^{(1)}(r-r_d) \\ N_{v'}^{(1)}(r-r_d) \end{bmatrix}, \quad (19)$$

where $a_v$ and $b_v$ are the expansion coefficients of the incident wave propagating along the $z$ direction, $R_{vv'}$ is the $2\times 2$ rotation block matrix, which is related to two sets of VSFs by the following relation

$$\begin{bmatrix} M_v^{(1,3)}(kr,\theta,\varphi) \\ N_v^{(1,3)}(kr,\theta,\varphi) \end{bmatrix} = \sum_{v'=0}^{\infty} R_{vv'}(\alpha,\beta,\gamma) \begin{bmatrix} M_{v'}^{(1,3)}(kr,\theta_1,\varphi_1) \\ N_{v'}^{(1,3)}(kr,\theta_1,\varphi_1) \end{bmatrix}, \quad (20)$$

where $r$, $\theta$, $\varphi$ and $r_1$, $\theta_1$, $\varphi_1$ are the spherical coordinates of the same evaluated point in the coordinate system $Oxyz$ and $Ox_1y_1z_1$, respectively. The coordinate system $Ox_1y_1z_1$ is obtained from the $Oxyz$ through the Euler rotation $(\alpha,\beta,\gamma)$. The EM field $(E_s, B_s)$ scattered by NPs in the absence of the molecule, are related to the incident wave by the T-Matrix, and they can be expressed as

$$E_s(\alpha,\beta,\gamma) = \sum_{v,v'',v'=1}^{\infty} [a_v \ b_v] R_{vv''}(-\gamma,-\beta,-\alpha) \sum_{\xi=1}^{NP} \begin{bmatrix} T^{11}_{v'v''}(\xi) & T^{12}_{v'v''}(\xi) \\ T^{21}_{v'v''}(\xi) & T^{22}_{v'v''}(\xi) \end{bmatrix}^T \begin{bmatrix} M^{(3)}_{v'}(r-r_\xi) \\ N^{(3)}_{v'}(r-r_\xi) \end{bmatrix}, \quad (21)$$

$$B_s(\alpha,\beta,\gamma) = [-i\sqrt{\varepsilon\varepsilon_0\mu\mu_0}] \sum_{v,v'',v'=1}^{\infty} [a_v \ b_v] R_{vv''}(-\gamma,-\beta,-\alpha)$$
$$\sum_{\xi=1}^{NP} \begin{bmatrix} T^{11}_{v'v''}(\xi) & T^{12}_{v'v''}(\xi) \\ T^{21}_{v'v''}(\xi) & T^{22}_{v'v''}(\xi) \end{bmatrix}^T \begin{bmatrix} N^{(3)}_{v'}(r-r_\xi) \\ M^{(3)}_{v'}(r-r_\xi) \end{bmatrix}, \quad (22)$$

where $\varepsilon$ ($\mu$) and $\varepsilon_0$ ($\mu_0$) are the relative and absolute permittivity (permeability) in the space and vacuum, respectively. $r_\xi$ is the coordinate of the $\xi^{th}$ sphere, $T^{ij}_{vv'}(\xi)$ are elements of the coupled T-matrix ($T(\xi)$) of the $\xi th$ NP, and they are related to the general single particle T-Matrix $\mathbf{T}_p$ by the following equations:

$$T(\xi) = \left( \sum_{p=1}^{NP} \mathscr{A}^{\xi p} \mathbf{T}_p S^{rt}_{p0} \right) \quad (23)$$

with $\mathscr{A} = A^{-1}$, and the block-matrix components of $A$ are written as $A^{\xi\xi} = I$ ($\xi = 1,2,...,NP$) and $A^{\xi p} = -\mathbf{T}_\xi \tilde{S}^{rtr}_{\xi p}$ ($\xi \neq p$). Here $\tilde{S}^{rtr}_{\xi p}$ and $S^{rt}_{p0}$ are transformation matrices for the coordinate systems defined in Appendix B of Ref. 34, and $I$ is the identity matrix. From Eqs (19-23), the absorption rate of the molecule can be expressed as

$$Q_{mol} = \frac{\omega_0 \gamma_{21}}{2} \frac{|\mu_{21} \cdot E' + m_{21} \cdot B'|^2}{|\hbar(\omega-\omega_0) + i\gamma_{21} - G|^2}$$
$$= \frac{1}{V} \sum_{\substack{v1,v1'',v1',\\v2,v2'',v2'=0}}^{\infty} [a_{v1} \ b_{v1}] R_{v1 v1''}(-\gamma,-\beta,-\alpha) \quad (24)$$
$$U_{v1''v1'} [U_{v2'v2''}]^\dagger [R_{v2v2''}(-\gamma,-\beta,-\alpha)]^\dagger \begin{bmatrix} a_{v2} \\ b_{v2} \end{bmatrix}^*$$

with

$$U_{v"v'} = \sum_{\xi=1}^{NP} \begin{bmatrix} T_{v'v"}^{11}(\xi) & T_{v'v"}^{12}(\xi) \\ T_{v'v"}^{21}(\xi) & T_{v'v"}^{22}(\xi) \end{bmatrix}^T \begin{bmatrix} \mathbf{M}_{v'}^{(3)}(\mathbf{r}_d - \mathbf{r}_\xi) \cdot \boldsymbol{\mu}_{21} - i\sqrt{\varepsilon\varepsilon_0\mu\mu_0} \mathbf{N}_{v'}^{(3)}(\mathbf{r}_d - \mathbf{r}_\xi) \cdot \mathbf{m}_{21} \\ \mathbf{N}_{v'}^{(3)}(\mathbf{r}_d - \mathbf{r}_\xi) \cdot \boldsymbol{\mu}_{21} - i\sqrt{\varepsilon\varepsilon_0\mu\mu_0} \mathbf{M}_{v'}^{(3)}(\mathbf{r}_d - \mathbf{r}_\xi) \cdot \mathbf{m}_{21} \end{bmatrix}$$
$$+ \begin{bmatrix} \delta_{v'v"} & 0 \\ 0 & \delta_{v'v"} \end{bmatrix} \begin{bmatrix} \mathbf{M}_{v'}^{(1)}(\mathbf{r}_d) \cdot \boldsymbol{\mu}_{21} - i\sqrt{\varepsilon\varepsilon_0\mu\mu_0} \mathbf{N}_{v'}^{(1)}(\mathbf{r}_d) \cdot \mathbf{m}_{21} \\ \mathbf{N}_{v'}^{(1)}(\mathbf{r}_d) \cdot \boldsymbol{\mu}_{21} - i\sqrt{\varepsilon\varepsilon_0\mu\mu_0} \mathbf{M}_{v'}^{(1)}(\mathbf{r}_d) \cdot \mathbf{m}_{21} \end{bmatrix} \quad (25)$$

Here $V = \omega_0 \gamma_{21} / (2|\hbar(\omega - \omega_0) + i\gamma_{21} - G|^2)$. The method to calculate the resonance broadening function $G$ has been described in Ref. 25, of which value is independent with the incident wave. From Eqs.(1-5) in the text, the absorption of NPs can be calculated. The total field around the $\xi$th NP is given by the following equation

$$\mathbf{E}_{tot}(\mathbf{r} - \mathbf{r}_\xi) = \mathbf{E}_{sct-\xi}(\mathbf{r} - \mathbf{r}_\xi) + \mathbf{E}_{\inf-\xi}(\mathbf{r} - \mathbf{r}_\xi), \quad (26)$$

where $\mathbf{E}_{sct-\xi}(\mathbf{r} - \mathbf{r}_\xi)$ is the total scattered field of the $\xi$th NP, which is expressed as

$$\mathbf{E}_{sct-\xi}(\mathbf{r} - \mathbf{r}_\xi) = \mathbf{E}_{s-\xi}^{(p)}(\mathbf{r} - \mathbf{r}_\xi) + \mathbf{E}_{s-\xi}^{(d)}(\mathbf{r} - \mathbf{r}_\xi). \quad (27)$$

Here $\mathbf{E}_{s-\xi}^{(p)}(\mathbf{r} - \mathbf{r}_\xi)$ and $\mathbf{E}_{s-\xi}^{(d)}(\mathbf{r} - \mathbf{r}_\xi)$ represent the scattered fields of the $\xi$th particle, which are caused by the external incident field and the irradiative molecule, respectively.

$$\mathbf{E}_{s-\xi}^{(x)}(\mathbf{r} - \mathbf{r}_\xi) = \sum_{v=1}^{\infty} \begin{bmatrix} c_v^{(x)}(\xi) & d_v^{(x)}(\xi) \end{bmatrix} \begin{bmatrix} \mathbf{M}_v^{(3)}(\mathbf{r} - \mathbf{r}_\xi) \\ \mathbf{N}_v^{(3)}(\mathbf{r} - \mathbf{r}_\xi) \end{bmatrix} \quad (x = 'd' \text{ or } 'p'), \quad (28)$$

where $c_v^{(x)}(\xi)$ and $d_v^{(x)}(\xi)$ are the expansion coefficients of the scattered wave, which are given by

$$\begin{bmatrix} c_v^{(p)}(\xi) \\ d_v^{(p)}(\xi) \end{bmatrix} = \sum_{v',v0=0}^{\infty} \begin{bmatrix} T_{vv'}^{11}(\xi) & T_{vv'}^{12}(\xi) \\ T_{vv'}^{21}(\xi) & T_{vv'}^{22}(\xi) \end{bmatrix} [R_{v0v'}(-\gamma, -\beta, -\alpha)]^T \begin{bmatrix} a_{v0} \\ b_{v0} \end{bmatrix} \quad (29)$$

and

$$\begin{bmatrix} c_v^{(d)}(\xi) \\ d_v^{(d)}(\xi) \end{bmatrix} = 2\sigma_{21} \begin{bmatrix} c_v^{(de)}(\xi) \\ d_v^{(de)}(\xi) \end{bmatrix} = \begin{bmatrix} \sum_{v0,v',v"=0}^{\infty} \frac{-1}{\Im} [a_{v0} \quad b_{v0}] R_{v0v'}(-\gamma, -\beta, -\alpha) U_{v'v"} \end{bmatrix} \begin{bmatrix} c_v^{(de)}(\xi) \\ d_v^{(de)}(\xi) \end{bmatrix} \quad (30)$$

with $\Im = \hbar(\omega - \omega_0) + i\gamma_{21} - G$. Here $c_v^{(de)}(\xi)$ and $d_v^{(de)}(\xi)$ are the expansion coefficients of the scattered wave from the $\xi^{th}$ NP irritated by a dipole with a momentum of $\boldsymbol{\mu}_{21}$, the calculated method for them can be found in the supplement materials of Ref. 25. Similarly,

$E_{\inf-\xi}(r-r_\xi) = E^{(p)}_{\inf-\xi}(r-r_\xi) + E^{(d)}_{\inf-\xi}(r-r_\xi)$ denotes the total incidental wave on the $\xi$th NP, which $E^{(p)}_{\inf-\xi}$ and $E^{(d)}_{\inf-\xi}$ are composed of the excited wave and the scattered fields from other NPs.

$$E^{(x)}_{\inf-\xi}(r-r_\xi) = \sum_{v=1}^{\infty} \begin{bmatrix} a^{(x)}_v(\xi) & b^{(x)}_v(\xi) \end{bmatrix} \begin{bmatrix} M^{(1)}_v(r-r_\xi) \\ N^{(1)}_v(r-r_\xi) \end{bmatrix} \quad (x = 'd' \text{ or } 'p') , \qquad (31)$$

where $a^{(x)}_v(\xi)$ and $b^{(x)}_v(\xi)$ are related to the scattering coefficients $c^{(x)}_{v'}(\xi)$ and $d^{(x)}_{v'}(\xi)$, that is

$$\begin{bmatrix} a^{(x)}_v(\xi) \\ b^{(x)}_v(\xi) \end{bmatrix} = \sum_{v,v'=1}^{\infty} \begin{bmatrix} [T^{-1}_\xi]^{11}_{vv'} & [T^{-1}_\xi]^{12}_{vv'} \\ [T^{-1}_\xi]^{21}_{vv'} & [T^{-1}_\xi]^{22}_{vv'} \end{bmatrix} \begin{bmatrix} c^{(x)}_{v'}(\xi) \\ d^{(x)}_{v'}(\xi) \end{bmatrix} . \qquad (32)$$

Using Eqs. (26-27), the absorption of NPs can be decomposed into the extinction part and the scattering part:

$$Q_{NP} = \sum_{\xi=1}^{NP} Q_{NP}(\xi) = \sum_{\xi=1}^{NP} (Q_{ext}(\xi) + Q_{sct}(\xi)) \qquad (33)$$

with $Q_{ext}(\xi) = \sum_{x,y=p,d} Q^{(x,y)}_{ext}(\xi)$ and $Q_{sct}(\xi) = \sum_{x,y=p,d} Q^{(x,y)}_{sct}(\xi)$, and

$$Q^{(x,y)}_{ext}(\xi) = -\frac{1}{2}\text{Re}\iint_{\Omega_\xi} n \cdot E^{(x)}_{\inf-\xi}(r-r_\xi) \times H^{(y)*}_{s-\xi}(r-r_\xi) + n \cdot E^{(y)}_{s-\xi}(r-r_\xi) \times H^{(x)*}_{\inf-\xi}(r-r_\xi) \, d\Omega ,$$

(34)

$$Q^{(x,y)}_{sct}(\xi) = -\frac{1}{2}\text{Re}\iint_{\Omega_\xi} n \cdot E^{(x)}_{s-\xi}(r-r_\xi) \times H^{(y)*}_{s-\xi}(r-r_\xi) \, d\Omega \qquad (35)$$

Using the orthogonality conditions of the VSFs, we have:

$$Q^{(x,y)}_{ext}(\xi) = -\frac{\pi}{2k^2}\sqrt{\frac{\varepsilon\varepsilon_0}{\mu\mu_0}}\sum_{v=1}^{\infty}\text{Re}\left\{\begin{bmatrix} c^{(x)}_v(\xi) & d^{(x)}_v(\xi) \end{bmatrix}\begin{bmatrix} a^{(y)*}_v(\xi) \\ b^{(y)*}_v(\xi) \end{bmatrix}\right\} \qquad (36)$$

and

$$Q^{(x,y)}_{sct}(\xi) = -\frac{\pi}{2k^2}\sqrt{\frac{\varepsilon\varepsilon_0}{\mu\mu_0}}\sum_{v=1}^{\infty}\text{Re}\left\{\begin{bmatrix} c^{(x)}_v(\xi) & d^{(x)}_v(\xi) \end{bmatrix}\begin{bmatrix} c^{(y)*}_v(\xi) \\ c^{(y)*}_v(\xi) \end{bmatrix}\right\} = Q^{(y,x)}_{sct}(\xi) . \qquad (37)$$

From Eqs. (26-37), we have

$$Q_{type}^{(x,y)}(\xi) = -\frac{\pi}{2k^2}\sqrt{\frac{\varepsilon\varepsilon_0}{\mu\mu_0}} \operatorname{Re}\left\{ \begin{bmatrix} z_{1\nu}^{(x)}(\xi) & z_{2\nu}^{(x)}(\xi) \end{bmatrix} \begin{bmatrix} s_{1\nu}^{(y)*}(\xi) \\ s_{2\nu}^{(y)*}(\xi) \end{bmatrix} \right\}$$

$$= -\frac{\pi}{2k^2}\sqrt{\frac{\varepsilon\varepsilon_0}{\mu\mu_0}} \sum_{\substack{v_0,v',\\v1_0 v1'=0}}^{\infty} \operatorname{Re}\left\{ \begin{bmatrix} a_{v_0}(\xi) & b_{v_0}(\xi) \end{bmatrix} \begin{bmatrix} R_{v_0 v'}(-\gamma,-\beta,-\alpha) \end{bmatrix} \right. \quad (38)$$

$$\left. \begin{bmatrix} L_{z\,v'v}^{(x)} \end{bmatrix} \begin{bmatrix} L_{s\,v1'v}^{(y)} \end{bmatrix}^{\dagger} \begin{bmatrix} R_{v1_0 v1'}(-\gamma,-\beta,-\alpha) \end{bmatrix}^{\dagger} \begin{bmatrix} a_{v1_0}(\xi) \\ b_{v1_0}(\xi) \end{bmatrix}^{*} \right\}$$

$$L_{z\,v'v}^{(x)} = \sum_{u=1}^{\infty} \mathrm{X}_{v'u}(x) \mathrm{P}_{uv}(z_1,z_2) \tag{39}$$

with

$$\mathrm{X}_{v'u}(x) = \begin{cases} \begin{bmatrix} T_{uv'}^{11}(\xi) & T_{uv'}^{12}(\xi) \\ T_{uv'}^{21}(\xi) & T_{uv'}^{22}(\xi) \end{bmatrix}^{T} & x \to p \\ \displaystyle\sum_{v''=1}^{\infty} \frac{-1}{\Im}[U_{v'v''}]\begin{bmatrix} c_u^{(de)}(\xi) & d_u^{(de)}(\xi) \end{bmatrix} & x \to d \end{cases}, \tag{40}$$

$$\mathrm{P}_{uv}(z_1,z_2) = \begin{cases} \begin{bmatrix} \delta_{uv} & 0 \\ 0 & \delta_{uv} \end{bmatrix} & (z_1,z_2 \to c,d) \\ \begin{bmatrix} [\mathbf{T}_l^{-1}]_{vu}^{11} & [\mathbf{T}_l^{-1}]_{vu}^{12} \\ [\mathbf{T}_l^{-1}]_{vu}^{21} & [\mathbf{T}_l^{-1}]_{vu}^{22} \end{bmatrix}^{T} & (z_1,z_2 \to a,b) \end{cases}. \tag{41}$$

Here when '*type*' is set to be '*ext*', $z_1,z_2 \to c,d$ and $s_1,s_2 \to a,b$, while $z_1,z_2 \to c,d$ and $s_1,s_2 \to c,d$ for '*sct*'. From the above equations, Eq.(6) in the text can be obtained.

**Symmetry analysis of** $\langle Q \rangle$. In this section, we present symmetry analysis of $\langle Q \rangle$. The expansion coefficients of any wave, $b_v$ and $a_v$, can also be expressed as [59-60]

$$b_v = \frac{1}{\sqrt{2n(n+1)\pi}} \frac{kr}{j_n(kr)} \int_{\theta=0}^{\pi} \int_{\varphi=0}^{2\pi} \mathbf{E}_0 \cdot e_r \bar{P}_n^m(\cos\theta) e^{-im\varphi} \sin\theta\, d\theta\, d\varphi, \tag{42}$$

$$a_v = -i\sqrt{\frac{\varepsilon\varepsilon_0\mu\mu_0}{2n(n+1)}} \frac{kr}{\pi j_n(kr)} \int_{\theta=0}^{\pi} \int_{\varphi=0}^{2\pi} \mathbf{B}_0 \cdot e_r \bar{P}_n^m(\cos\theta) e^{-im\varphi} \sin\theta\, d\theta\, d\varphi, \tag{43}$$

where $\bar{P}_n^m(\cos\theta)$ is related to the associated Legendre function, which is given by

$$\overline{P}_n^m(\cos\theta) = \sqrt{\frac{2n+1}{2}\frac{(n-m)!}{(n+m)!}} P_n^m(\cos\theta). \tag{44}$$

If the space inversions are performed for both the NP-molecule system and the incident wave, the absorption rate of the system $\langle \tilde{Q}_{mol/NP} \rangle$ should be unchanged because of parity conservation[61]. Since the electric field possesses an odd parity while the magnetic field has an even parity, from Eqs.(42) and (43), we have

$$\tilde{a}_v = (-1)^{n+1} a_v; \quad \tilde{b}_v = (-1)^n b_v, \tag{45}$$

where $\tilde{a}_v$ and $\tilde{b}_n$ are expansion coefficients of the space inversed incident wave, while $\langle \tilde{M}_{mol/NP} \rangle_\Omega$ is determined only by the property of the enantiomorphic system. According to Eq.(10) in the main text, we have

$$\begin{aligned}
\langle Q_{mol/NP} \rangle &= \sum_{v=1}^{\infty} \mathrm{Re}\left\{\left[\langle M_{mol/NP} \rangle_\Omega\right]_{vv}^{11} |a_v|^2 + \left[\langle M_{mol/NP} \rangle_\Omega\right]_{vv}^{22} |b_v|^2 \right. \\
&\left. + \left[\langle M_{mol/NP} \rangle_\Omega\right]_{vv}^{12} a_v b_v^* + \left[\langle M_{mol/NP} \rangle_\Omega\right]_{vv}^{21} b_v a_v^* \right\} \\
&= \sum_{v=1}^{\infty} \mathrm{Re}\left\{\left[\langle \tilde{M}_{mol/NP} \rangle_\Omega\right]_{vv}^{11} |\tilde{a}_v|^2 + \left[\langle \tilde{M}_{mol/NP} \rangle_\Omega\right]_{vv}^{22} |\tilde{b}_v|^2 \right. \\
&\left. + \left[\langle \tilde{M}_{mol/NP} \rangle_\Omega\right]_{vv}^{12} \tilde{a}_v \tilde{b}_v^* + \left[\langle \tilde{M}_{mol/NP} \rangle_\Omega\right]_{vv}^{21} \tilde{b}_v \tilde{a}_v^* \right\}
\end{aligned} \tag{46}$$

Form Eqs (45-46), it can be seen that $\left[\langle M_{mol/NP} \rangle_\Omega\right]_{vv}^{12}$ and $\left[\langle M_{mol/NP} \rangle_\Omega\right]_{vv}^{21}$ are connected with $\left[\langle \tilde{M}_{mol/NP} \rangle_\Omega\right]_{vv}^{12}$ and $\left[\langle \tilde{M}_{mol/NP} \rangle_\Omega\right]_{vv}^{21}$ through

$$\mathrm{Re}\left[\left[\langle M_{mol/NP} \rangle_\Omega\right]_{vv}^{12} + \left[\langle M_{mol/NP} \rangle_\Omega\right]_{vv}^{21}\right] = -\mathrm{Re}\left[\left[\langle \tilde{M}_{mol/NP} \rangle_\Omega\right]_{vv}^{12} + \left[\langle \tilde{M}_{mol/NP} \rangle_\Omega\right]_{vv}^{21}\right] \tag{47}$$

and

$$\mathrm{Im}\left[\left[\langle M_{mol/NP} \rangle_\Omega\right]_{vv}^{12} - \left[\langle M_{mol/NP} \rangle_\Omega\right]_{vv}^{21}\right] = -\mathrm{Im}\left[\left[\langle \tilde{M}_{mol/NP} \rangle_\Omega\right]_{vv}^{12} - \left[\langle \tilde{M}_{mol/NP} \rangle_\Omega\right]_{vv}^{21}\right] \tag{48}$$

At the same time $\left[\langle M_{mol/NP} \rangle_\Omega\right]_{vv}^{11}$ and $\left[\langle M_{mol/NP} \rangle_\Omega\right]_{vv}^{22}$ should satisfy

$$\mathrm{Re}\left[\langle M_{mol/NP} \rangle_\Omega\right]_{vv}^{11} = \mathrm{Re}\left[\langle \tilde{M}_{mol/NP} \rangle_\Omega\right]_{vv}^{11} \tag{49}$$

and

$$\text{Re}\left[\langle M_{mol/NP}\rangle_\Omega\right]_{vv}^{22} = \text{Re}\left[\langle \tilde{M}_{mol/NP}\rangle_\Omega\right]_{vv}^{22}, \tag{50}$$

if Eq. (46) will be satisfied for all kinds of incident waves.

According to the equations above, the dichroism of the system can be written as

$$\begin{aligned}\Delta &= \boldsymbol{a}\langle M_{mol/NP}\rangle_\Omega \boldsymbol{a}^\dagger - \tilde{\boldsymbol{a}}\langle M_{mol/NP}\rangle_\Omega \tilde{\boldsymbol{a}}^\dagger \\ &= 2\sum_{v=1}^\infty \text{Re}\left\{\left[\langle M_{mol/NP}\rangle_\Omega\right]_{vv}^{12} a_v b_v^* + \left[\langle M_{mol/NP}\rangle_\Omega\right]_{vv}^{21} b_v a_v^*\right\}\end{aligned}. \tag{51}$$

For the special case of the OAM beams, Eqs.(10-11) in the text can be readily gotten.

**Figure Captions**

Figure 1 | (a) System of coordinates and schematics of a complex composed of a gold nanoparticle and chiral molecule. (b) and (c) Orientation averaged *OD* as a function of wavelength for a metal nanoparticle and chiral molecule under the OAM incident beam with $l = \pm 1$ and $l = \pm 2$, respectively. The insets represent the calculated results for the single chiral molecule without NPs. (d) System of coordinates and schematics of a NP dimer and a chiral molecule. (e) and (f) Calculated OD as a function of wavelengths for the Au dimer and a chiral molecule under the OAM incident beam with $l = \pm 1$ and $l = \pm 2$, respectively. Calculated extinctions for the corresponding systems are also shown. The radius of NPs are taken 15nm and the separation between two NPs is 1nm.

Figure 2 | Orientation averaged OD as a function of wavelength under the OAM incident beam when the nanocomposites locate at $x = 50nm$ (olive line), $100nm$ (red line) and $150nm$ (blue dark line), respectively. The corresponding signals for the mirror reflected system are presented as green ( $x = 50nm$ ), pink ( $x = 100nm$ ), blue ( $x = 150nm$ ). (a) and (b) correspond to the system of a metal nanoparticle and chiral molecule with $l = \pm 1$ and $l = \pm 2$, respectively. (c) and (d) to the system of a NP dimer and a chiral molecule with $l = \pm 1$ and $l = \pm 2$, respectively. (e) and (f) describe $\mathrm{Im}[\boldsymbol{E}_{inc+}^{(+l)} \cdot \boldsymbol{H}_{inc+}^{(+l)*}] / |\boldsymbol{E}_{inc+}^{(+l)}|^2$ as a function of position for the incident OAM beam with $l = 1$ and $l = 2$ at the wavelength of 300nm, respectively.

Figure3 | (a) The OD distribution in the focal plane at the wavelength of 300nm for the

molecule-NP system described in Fig.1a. The OAM beams have $NA = 0.55$ and $l = \pm 1$. (b) The corresponding intensity distribution of the electric field at the wavelength of 300nm.

Figure 4 | (a) The distribution of nanocomposites with fixed orientations in a circular region. (b) Schematics of the OAM beam and the regions being considered. The green region marks ρ=0-50nm, red region corresponds to ρ=50-100nm, blue region to ρ=100-150nm. (c) and (d) Orientation fixed OD averaged in some regions as a function of wavelengths for a metal nanoparticle and a chiral molecule under the OAM incident beam with $l = \pm 1$ and $l = \pm 2$, respectively. The insets represent the corresponding results for the single chiral molecule without NPs. (e) and (f) Orientation fixed OD averaged in some regions as a function of wavelengths for the Au dimer and a chiral molecule under the OAM incident beam with $l = \pm 1$ and $l = \pm 2$, respectively. The olive/red/dark blue lines correspond to the signals in ρ=0-50nm/ ρ=50-100nm/ ρ=100-150nm regions. The corresponding signals for the mirror reflected system are presented as green/pink/blue lines. (g) and (h) Averaged relative values of the longitude electric field magnitude as a function of the relative distances between the center of the x-polarized OAM beam and nanocomposites for the case with $l = 1$ and $l = 2$, respectively. Panel A: electric field distribution for a dimmer when the molecule is put at $x = 100nm$ in the focal plane; Panel B: electric field distribution for a dimmer when the molecule is put at the center of the beam. The other parameters are taken identical with those in Fig. 1.

Figure 5 | (a) and (b) Orientation averaged OD in some regions as a function of wavelength

for a metal nanoparticle and chiral molecule under the OAM incident beam with $l=\pm 1$ and $l=\pm 2$, respectively. The insets represent the corresponding results for the single chiral molecule without NPs. (c) and (d) Orientation averaged OD in some regions as a function of wavelength for the Au dimer and a chiral molecule under the OAM incident beam with $l=\pm 1$ and $l=\pm 2$, respectively. The olive/red/dark blue lines correspond to the signals in ρ=0-50nm/ ρ=50-100nm/ ρ=100-150nm regions. The corresponding signals for the mirror reflected system are presented as green/pink/blue lines.

**AUTHOR INFORMATION**

**Author contributions**

Numerical results and theoretical method are presented by T. W., the idea and physical analysis are given by X. Z. and R.W. All authors reviewed the manuscript.

**Competing financial interests:**

The authors declare no competing financial interest.

**Acknowledgment**

This work was supported by the National Natural Science Foundation of China (Grant No. 11274042 , 61421001 and 11174033).

**Additional information:**

Supplementary information accompanies this paper at

http://www.nature.com/scientificreports

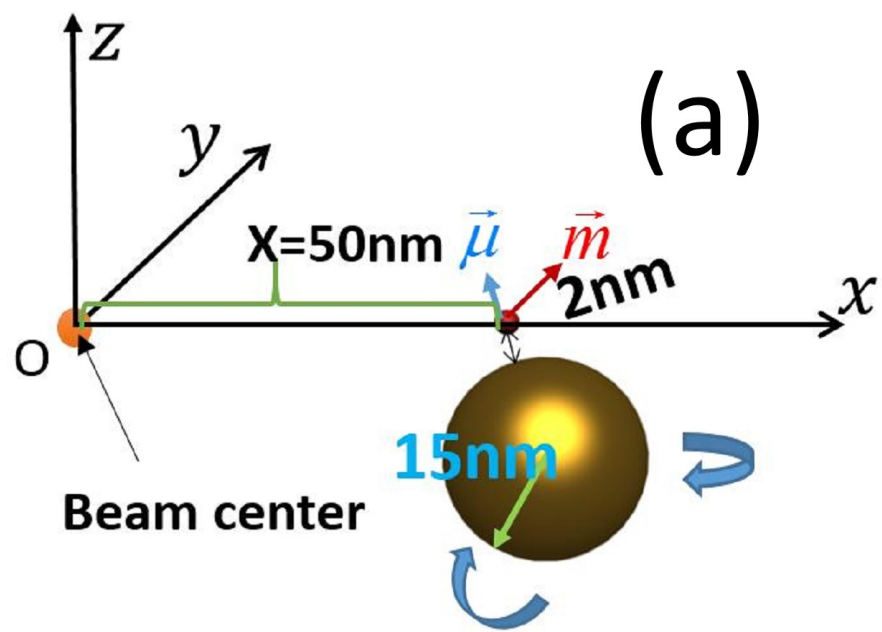
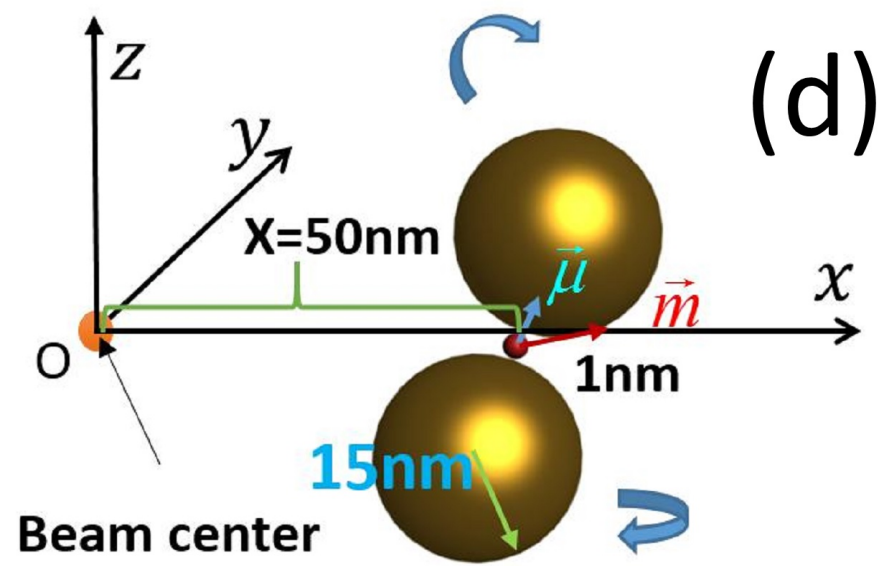
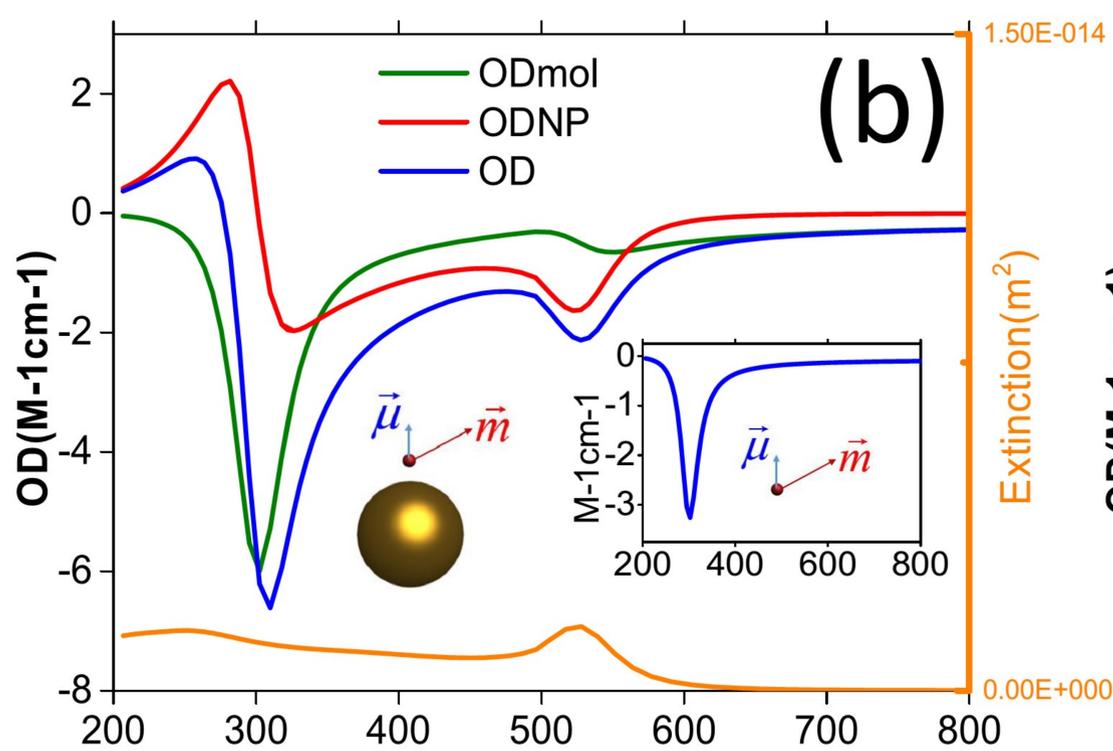
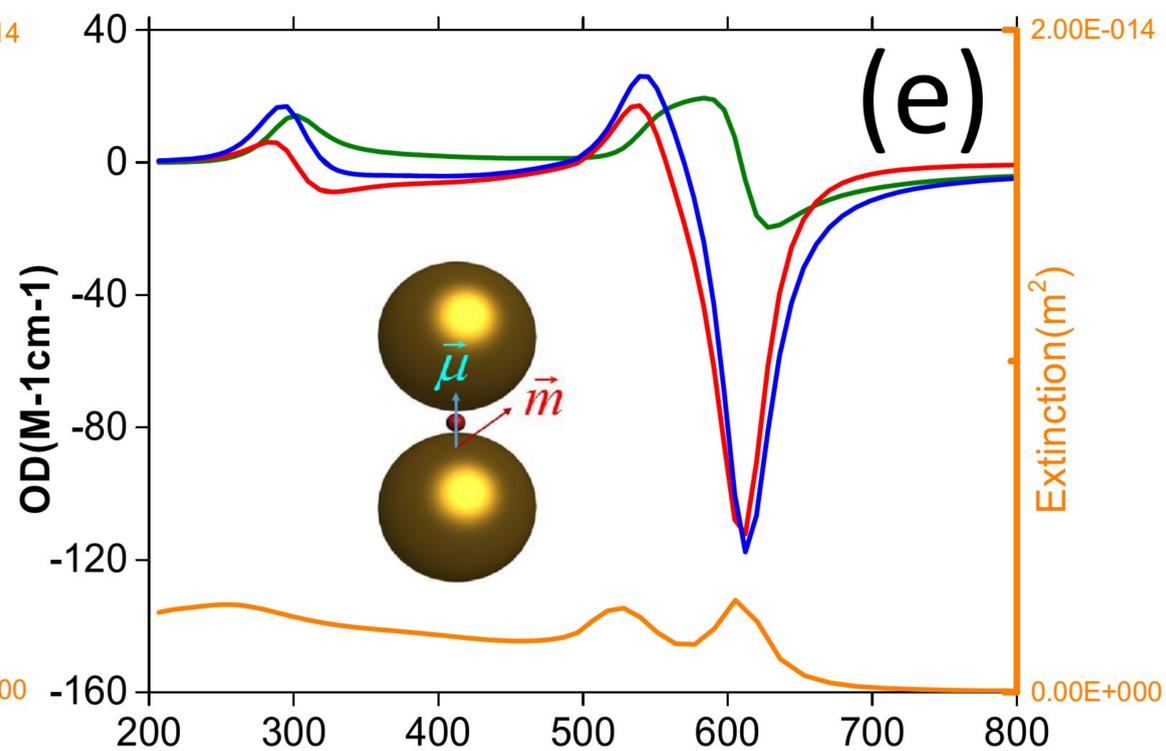
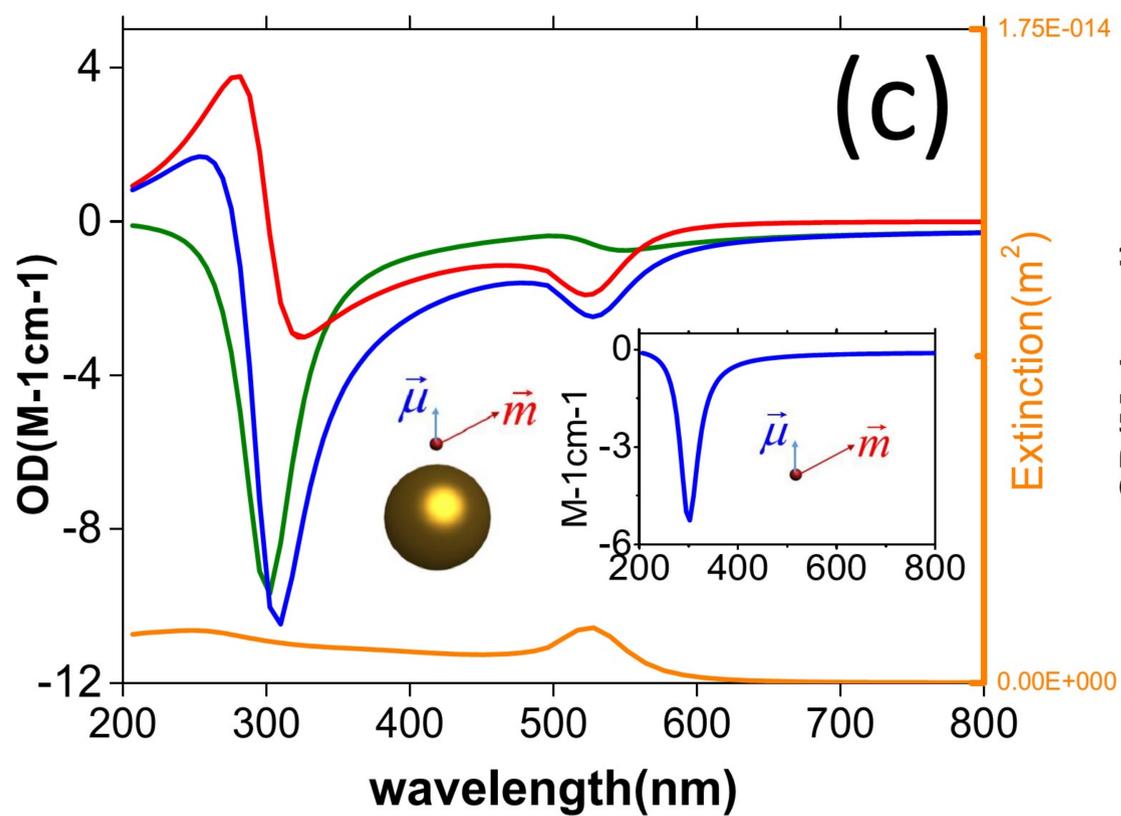
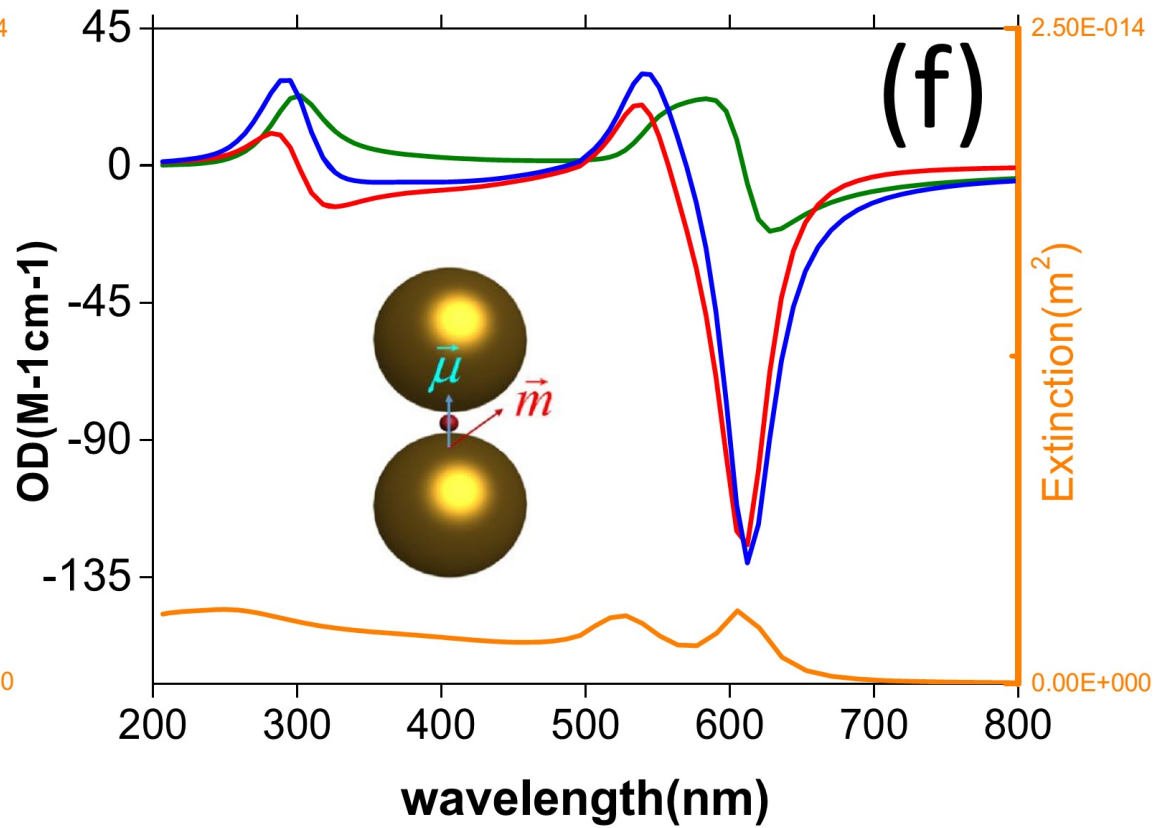

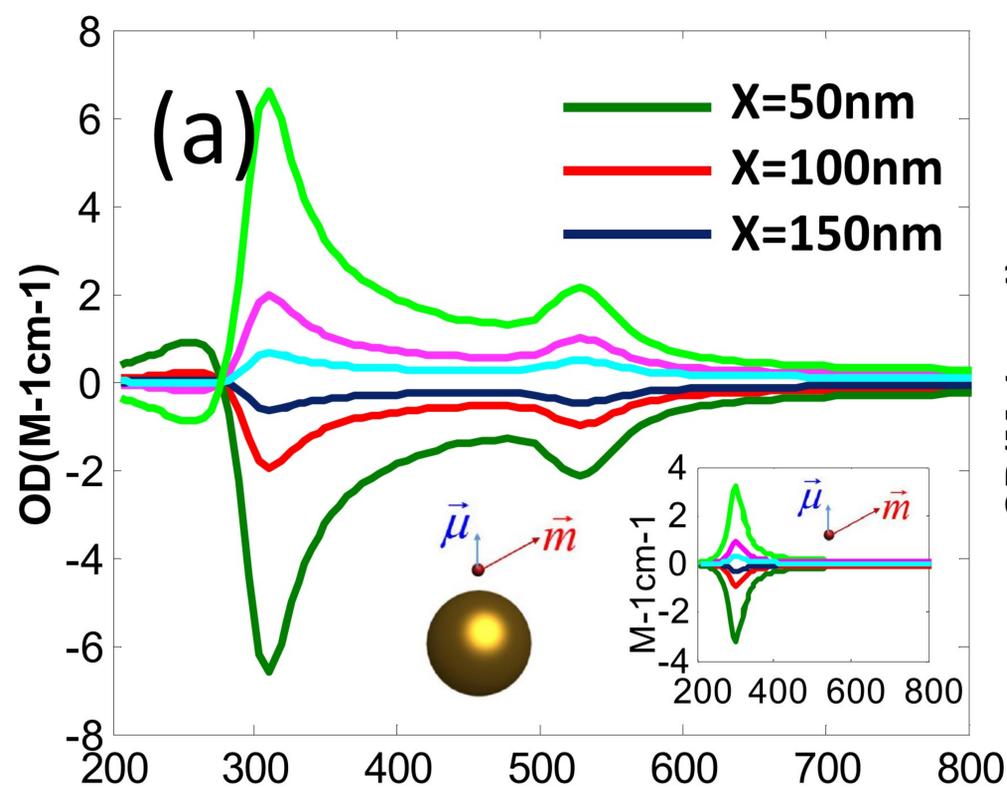
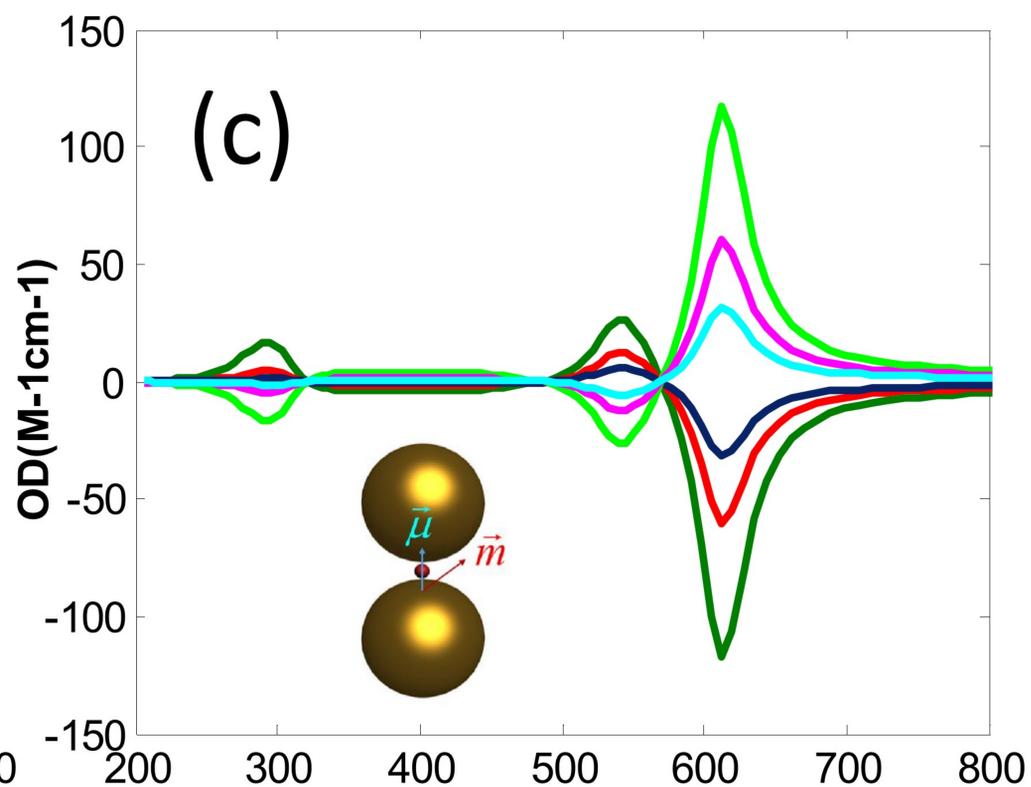
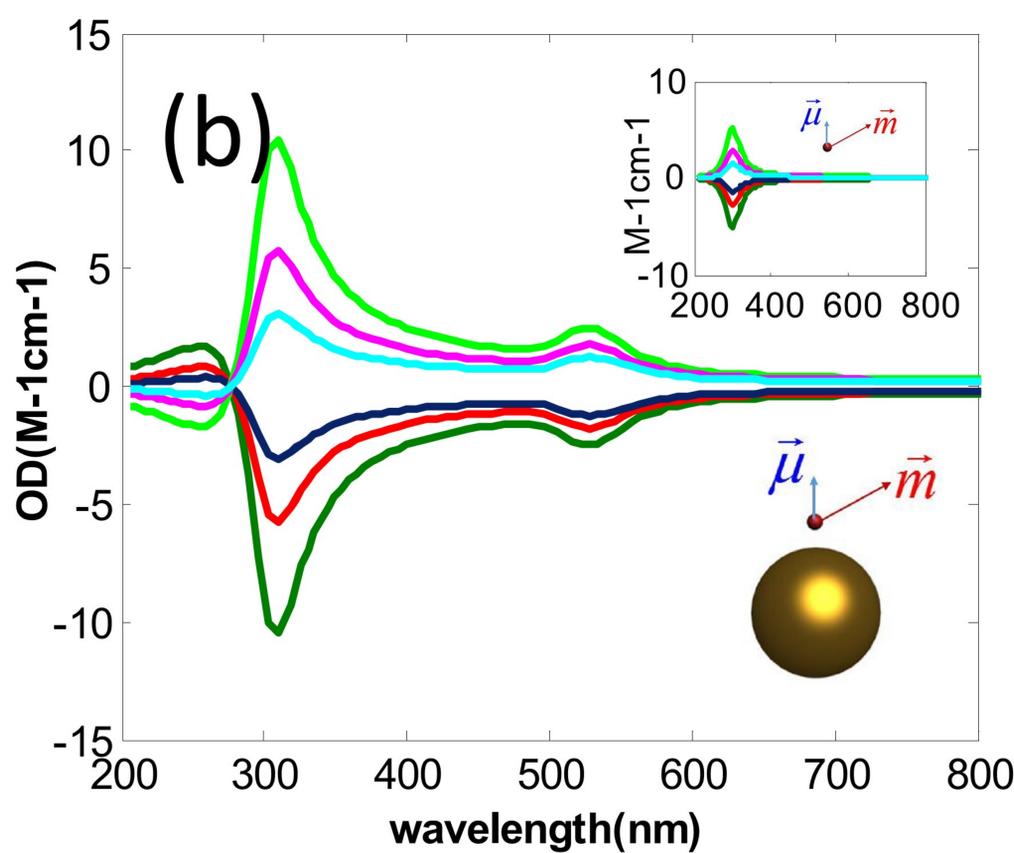
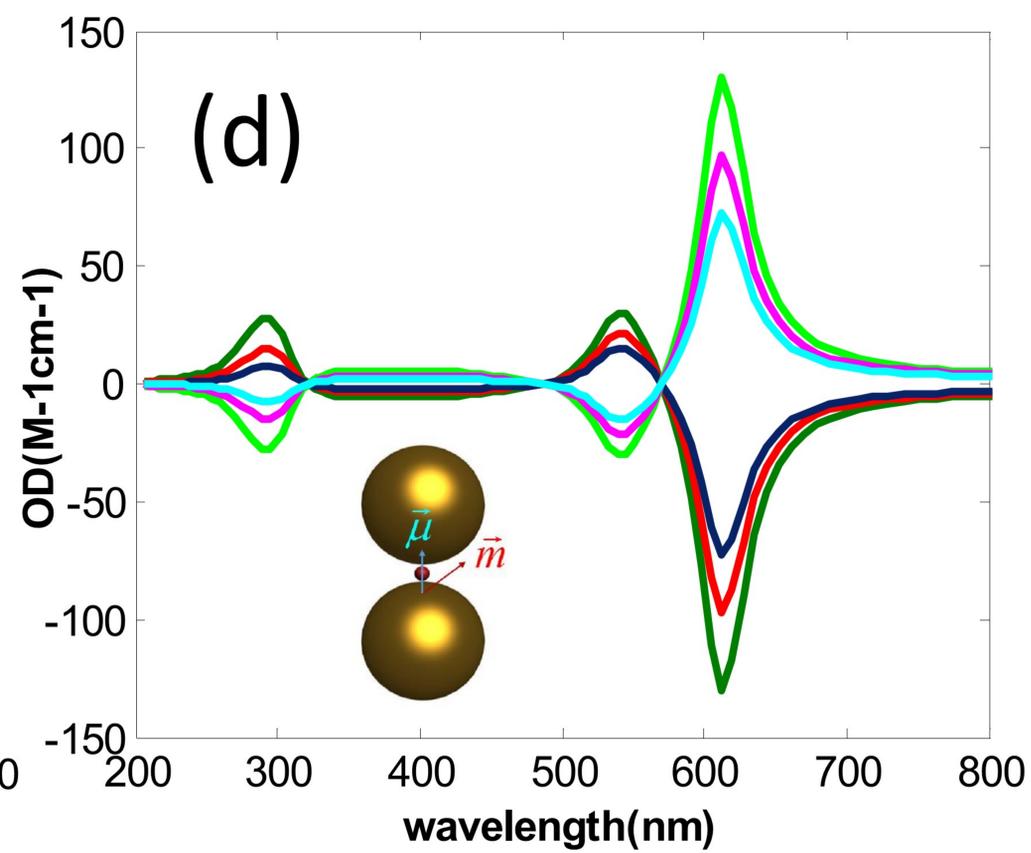
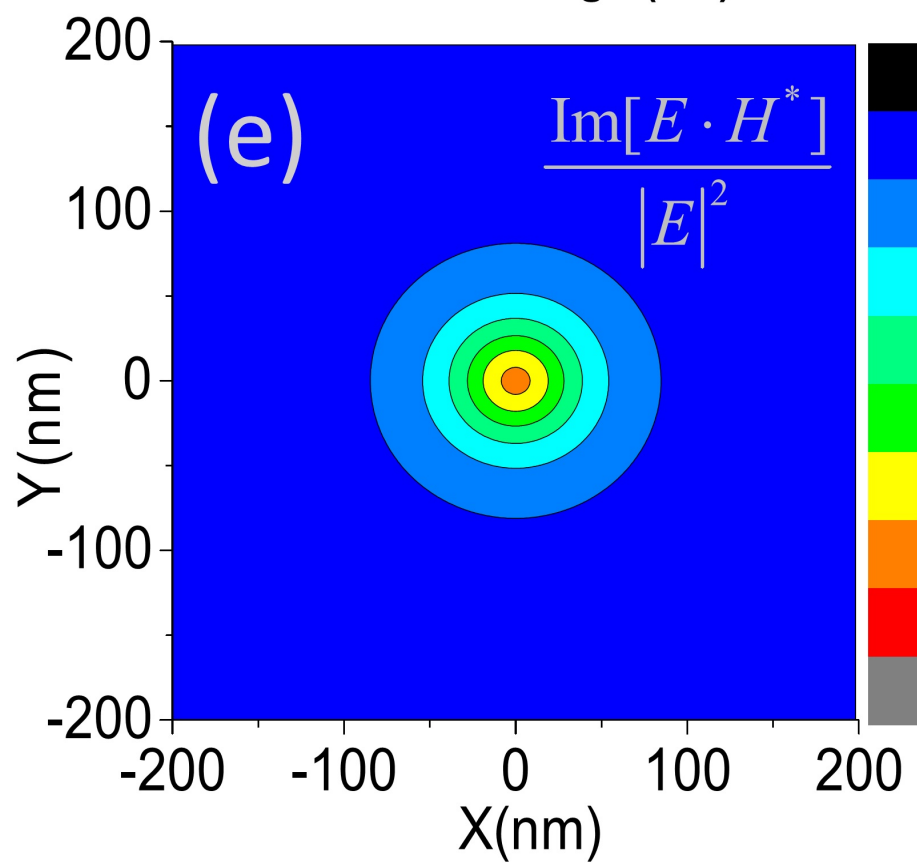
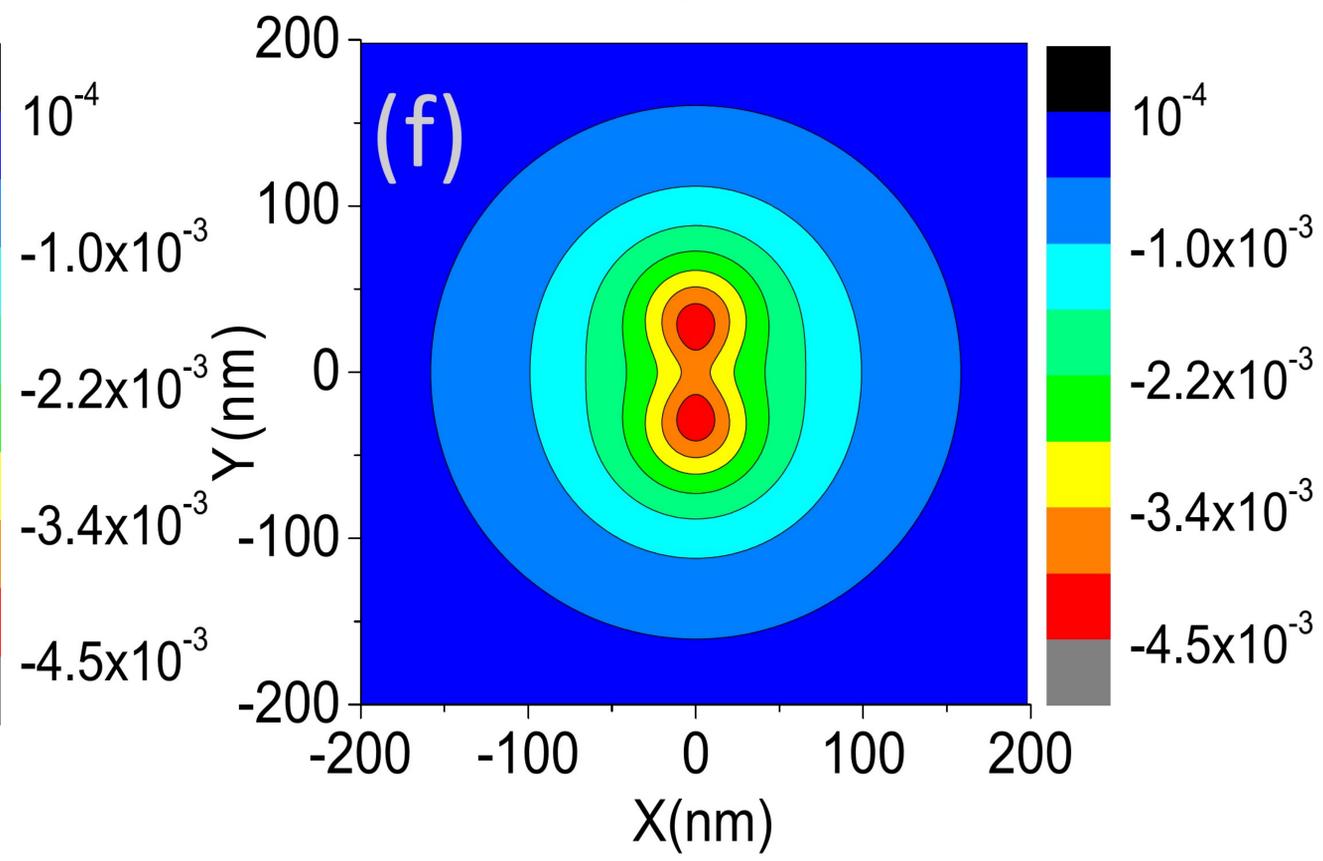

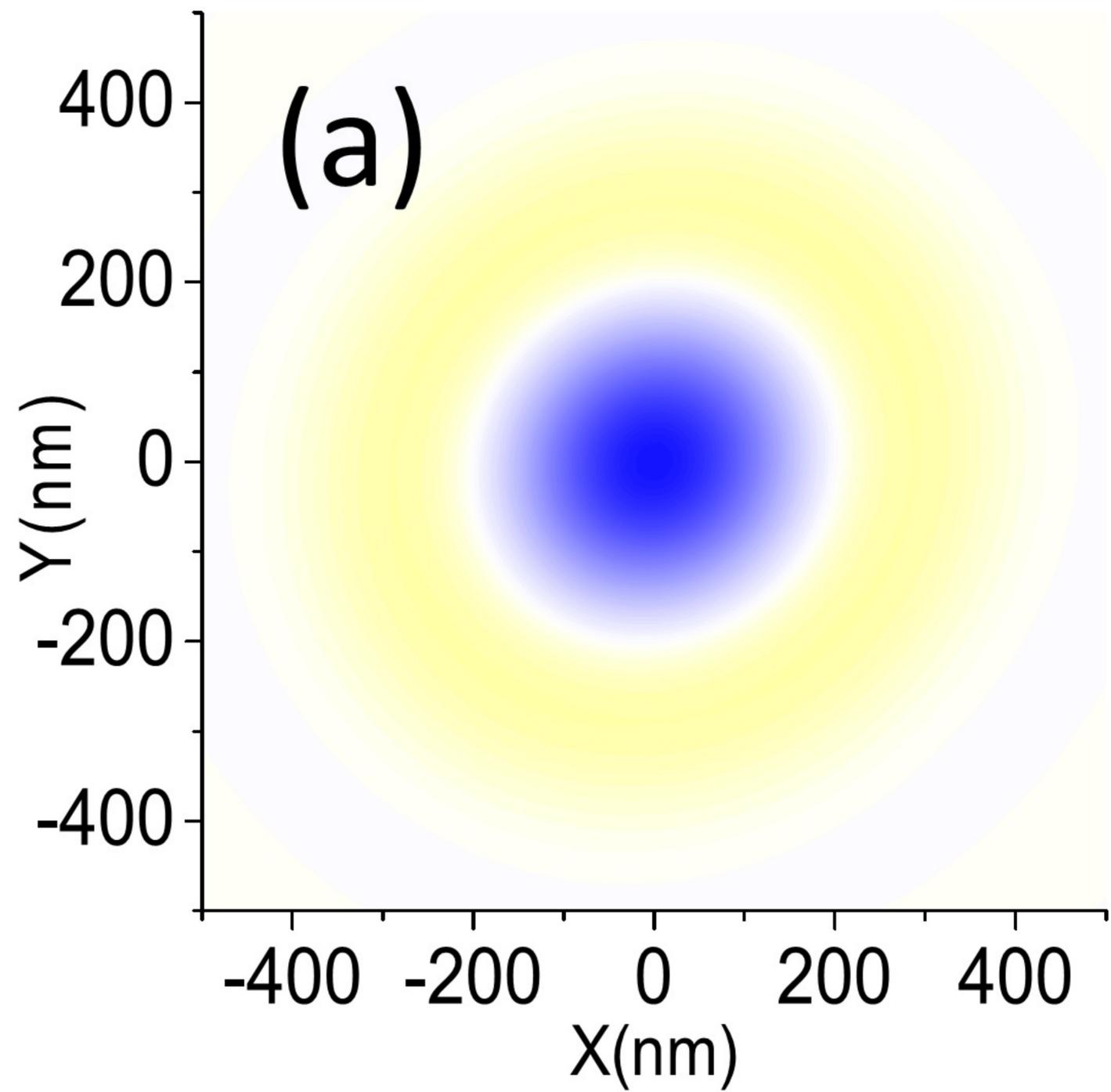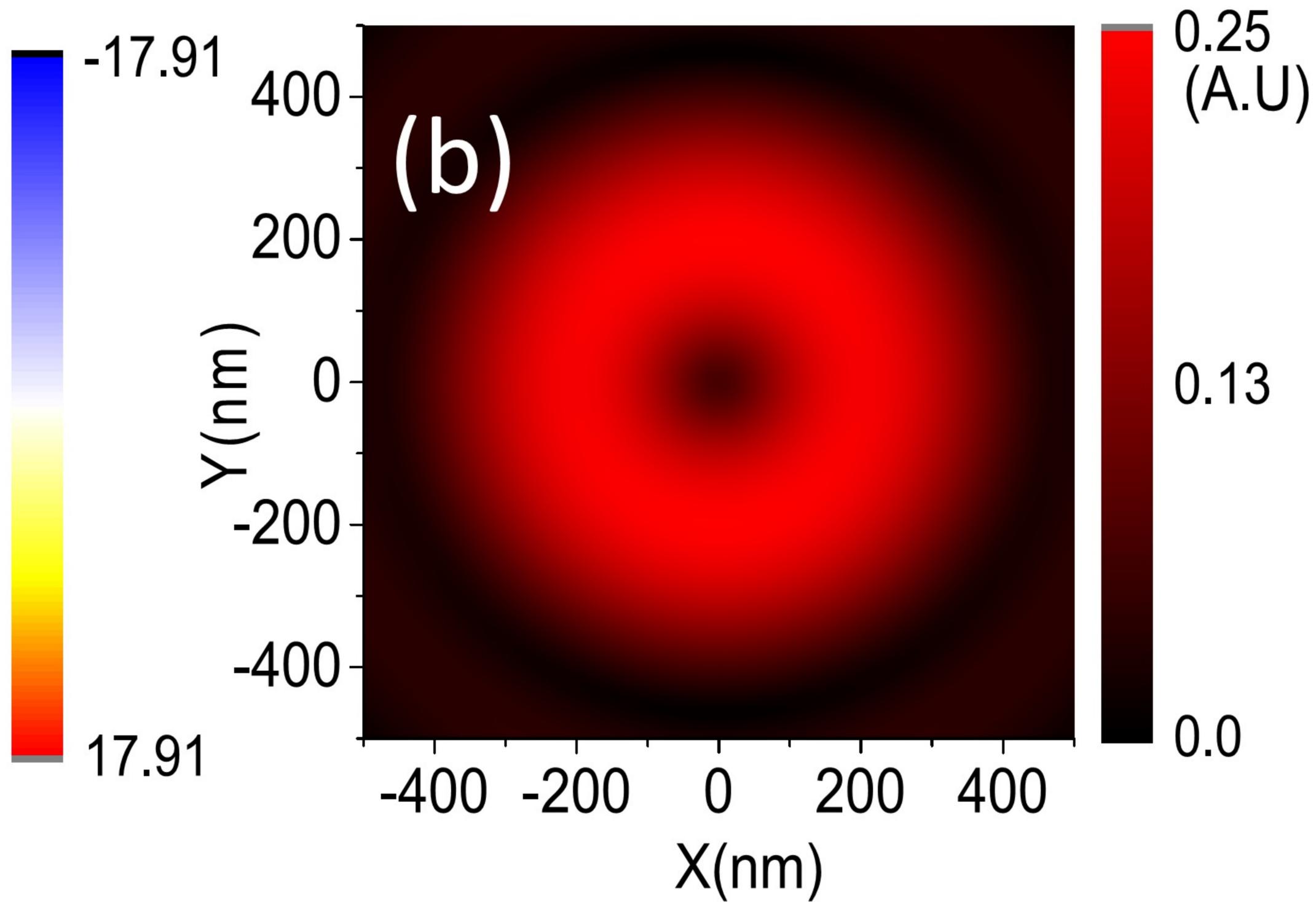

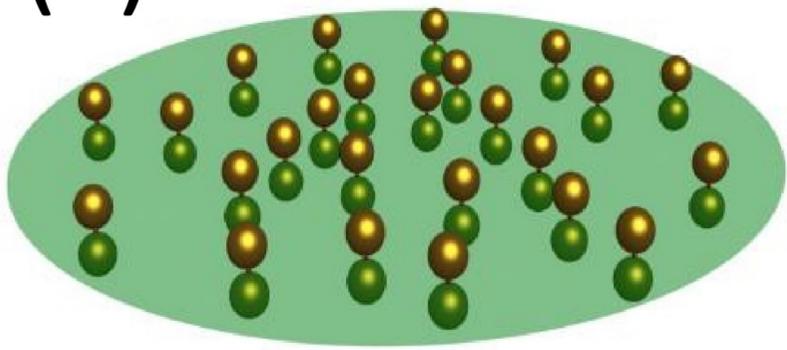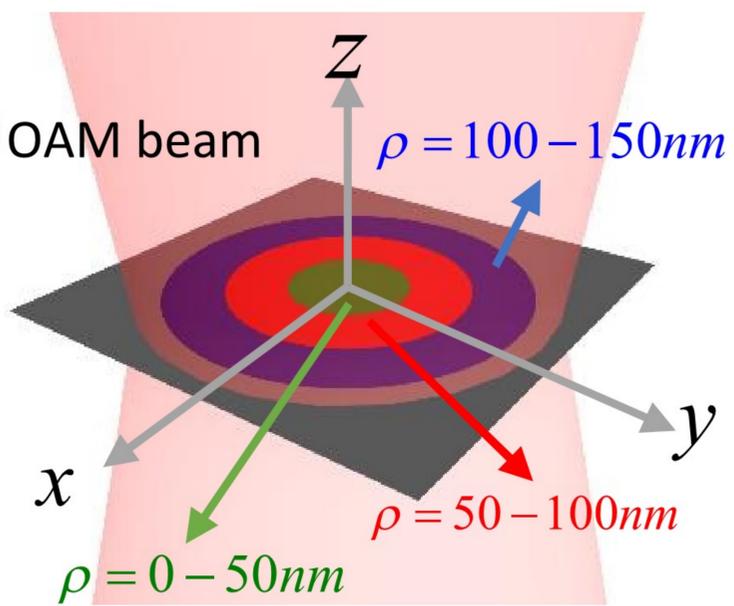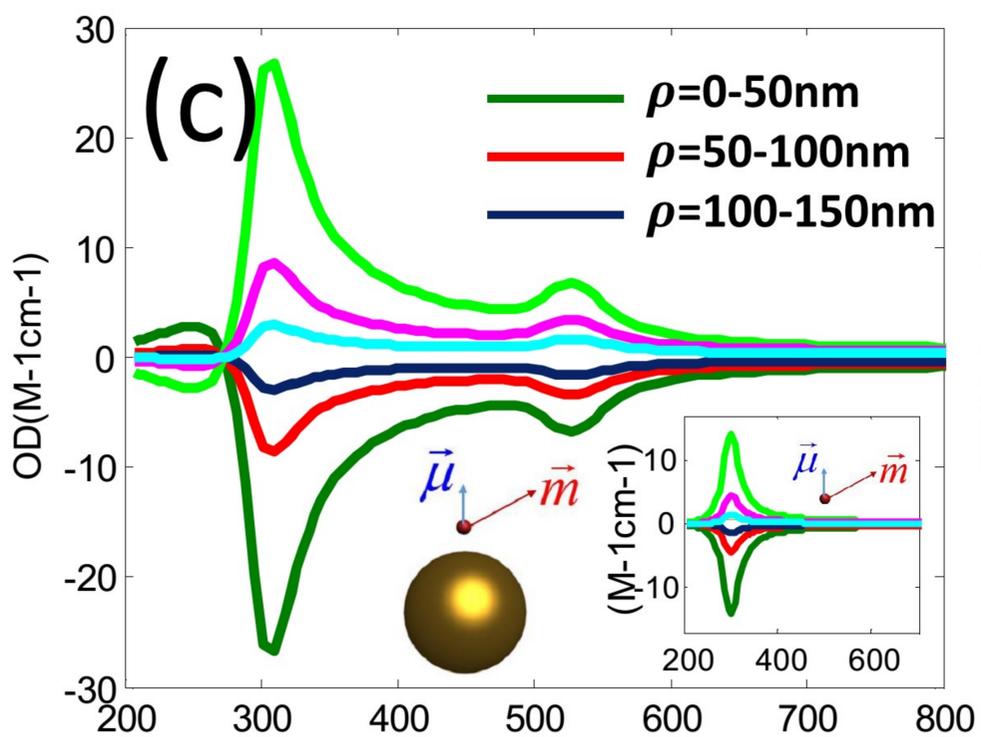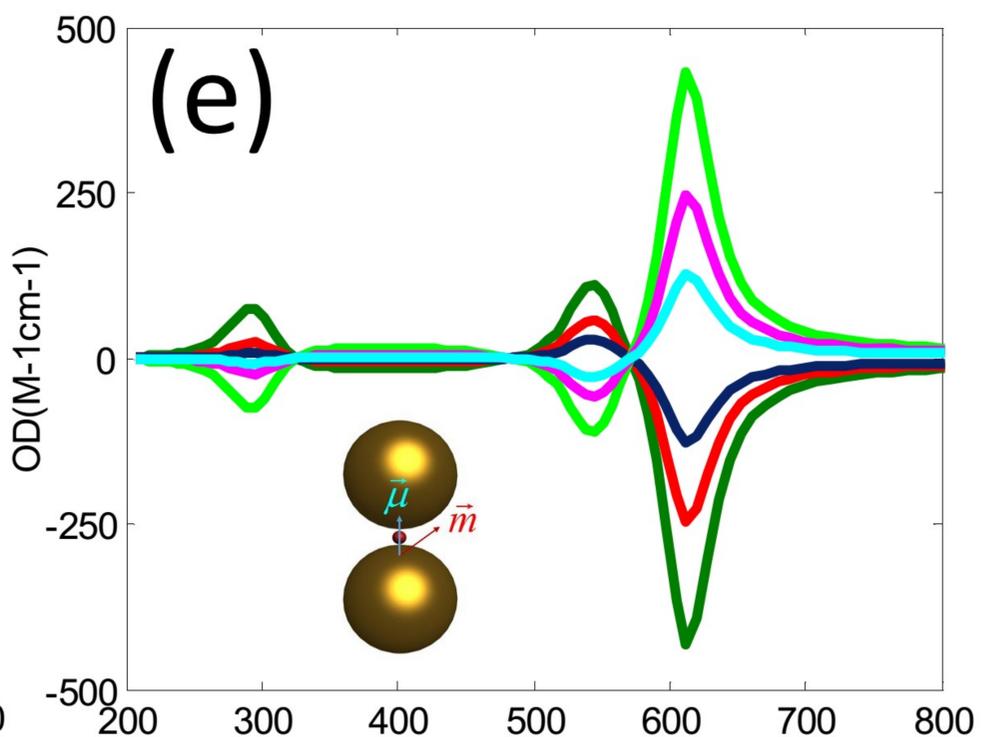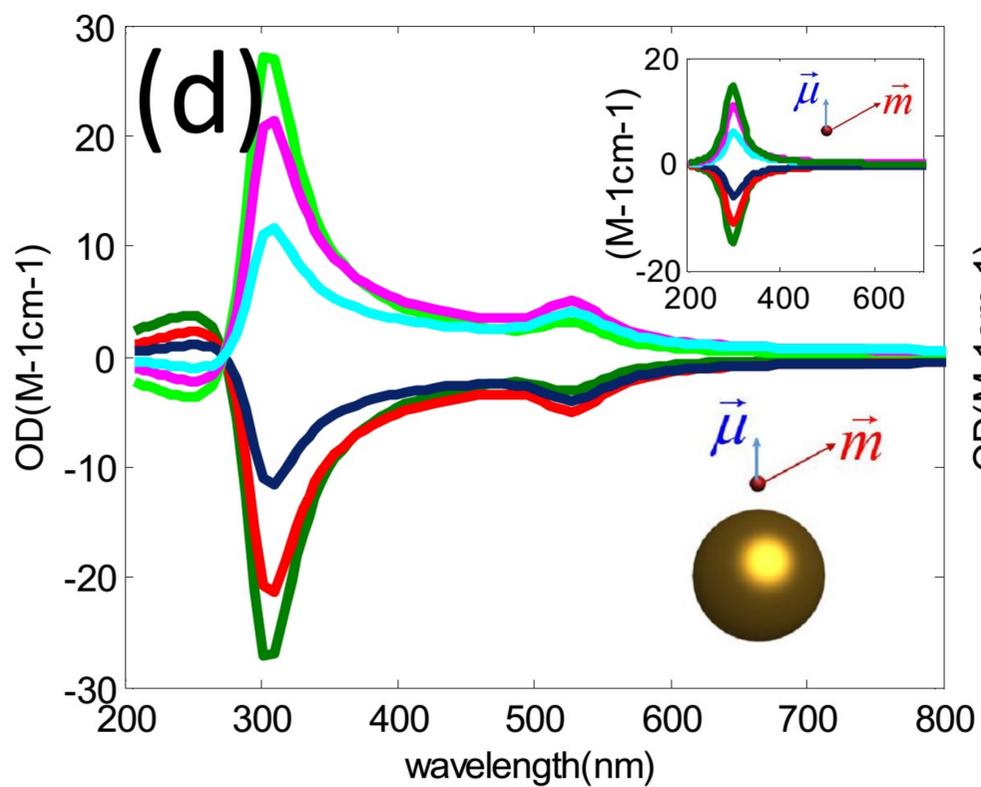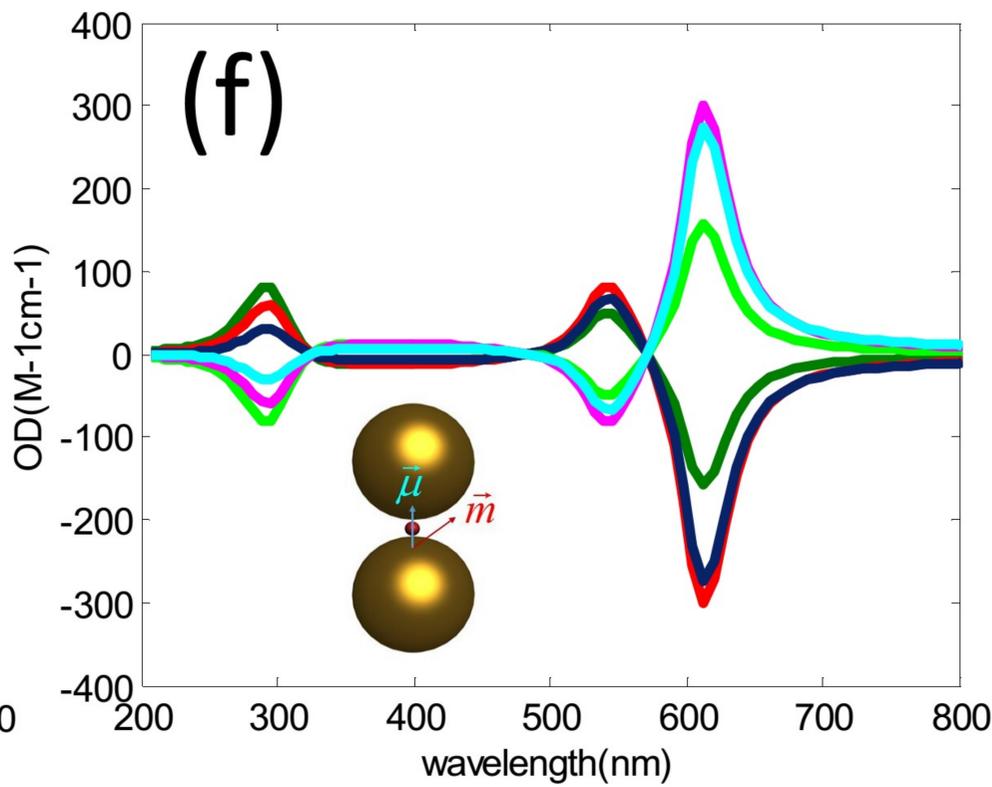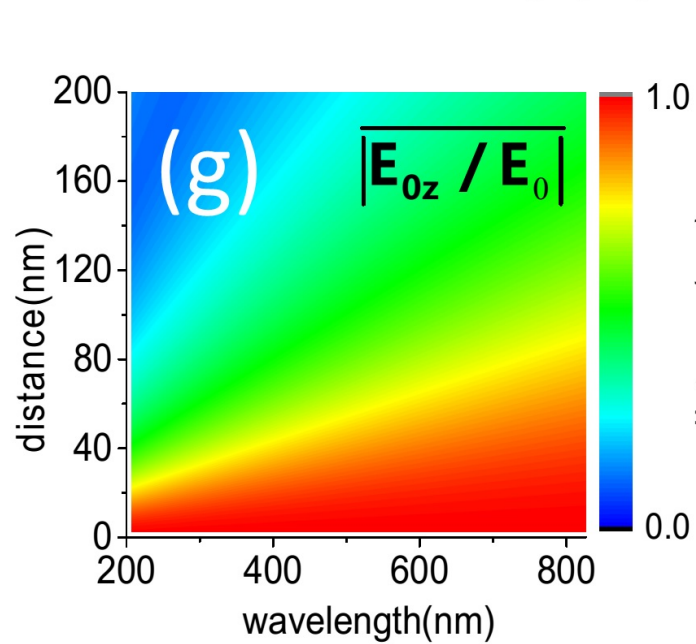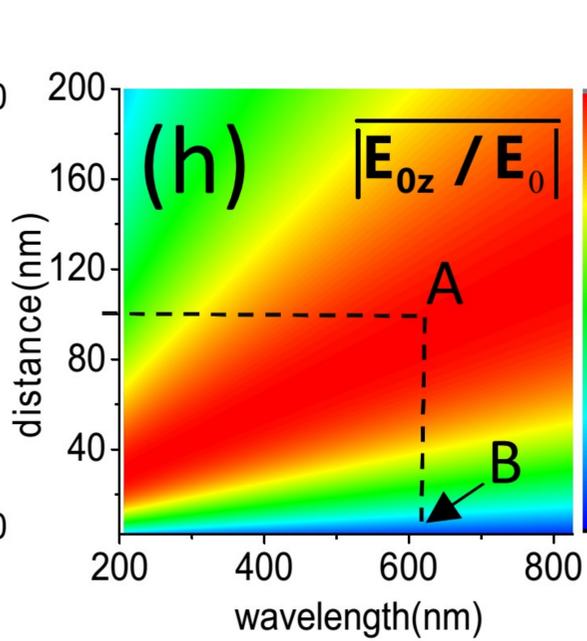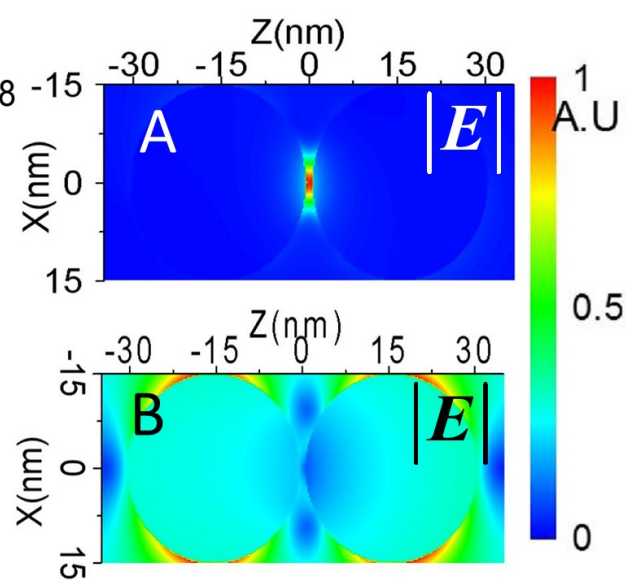

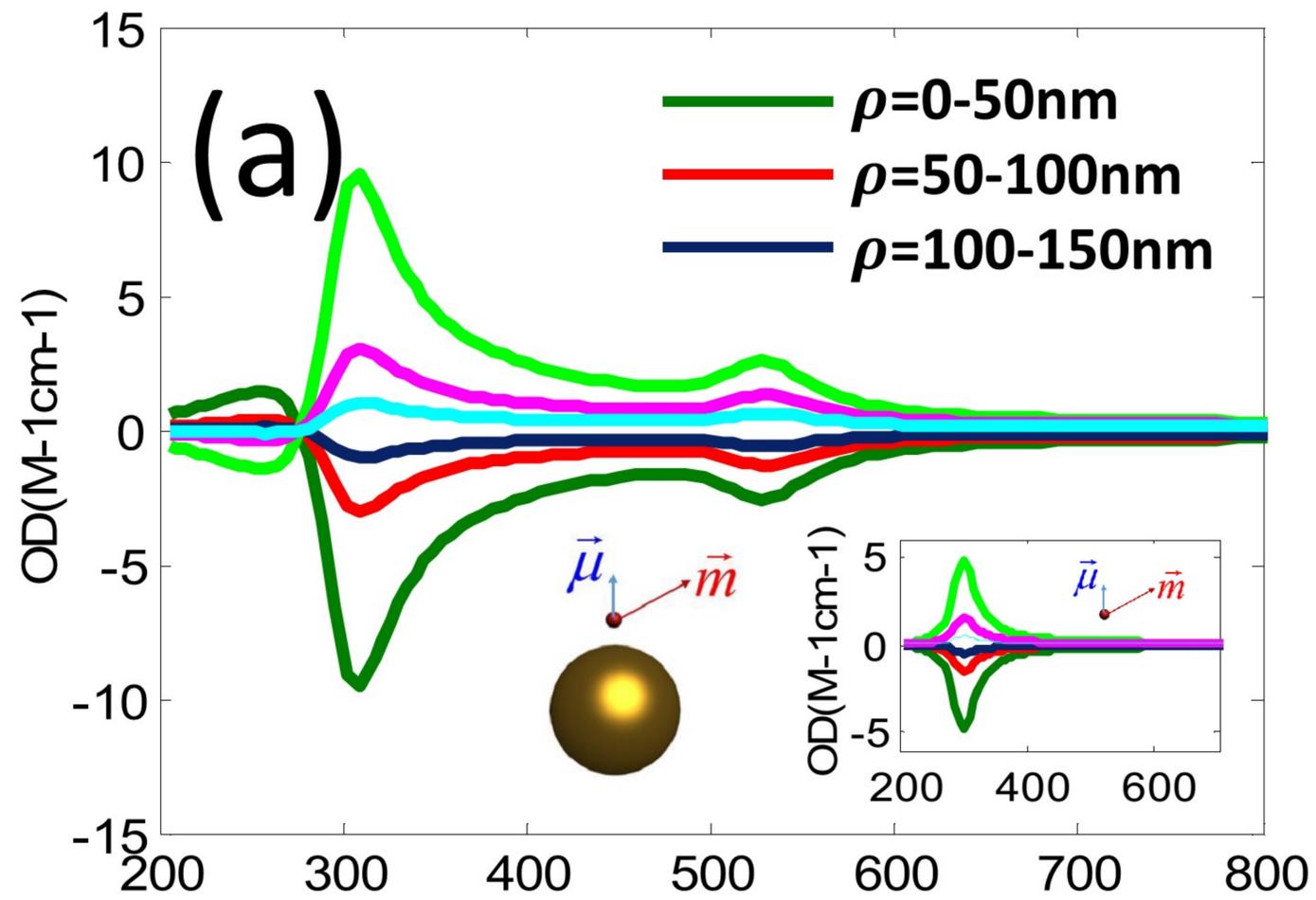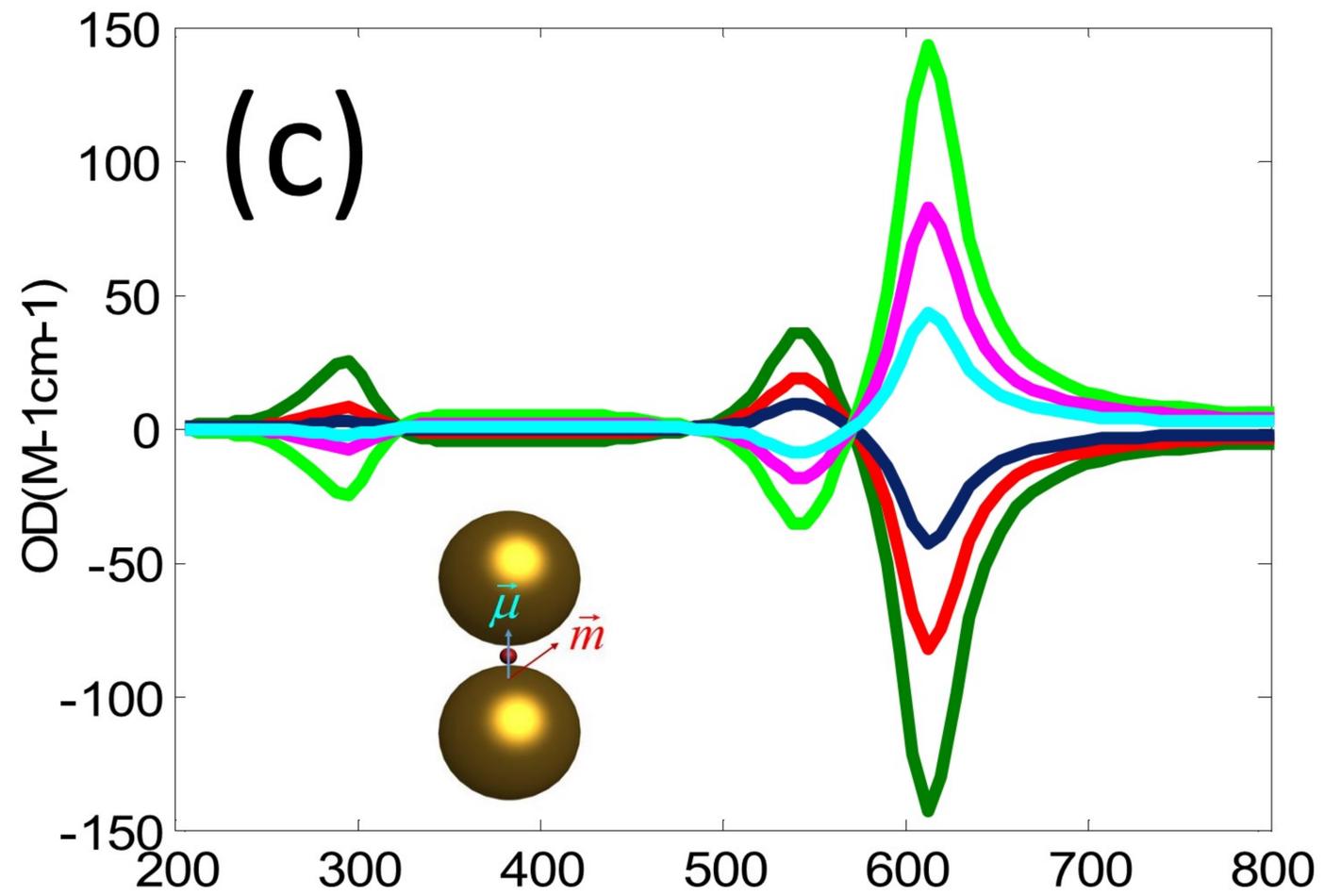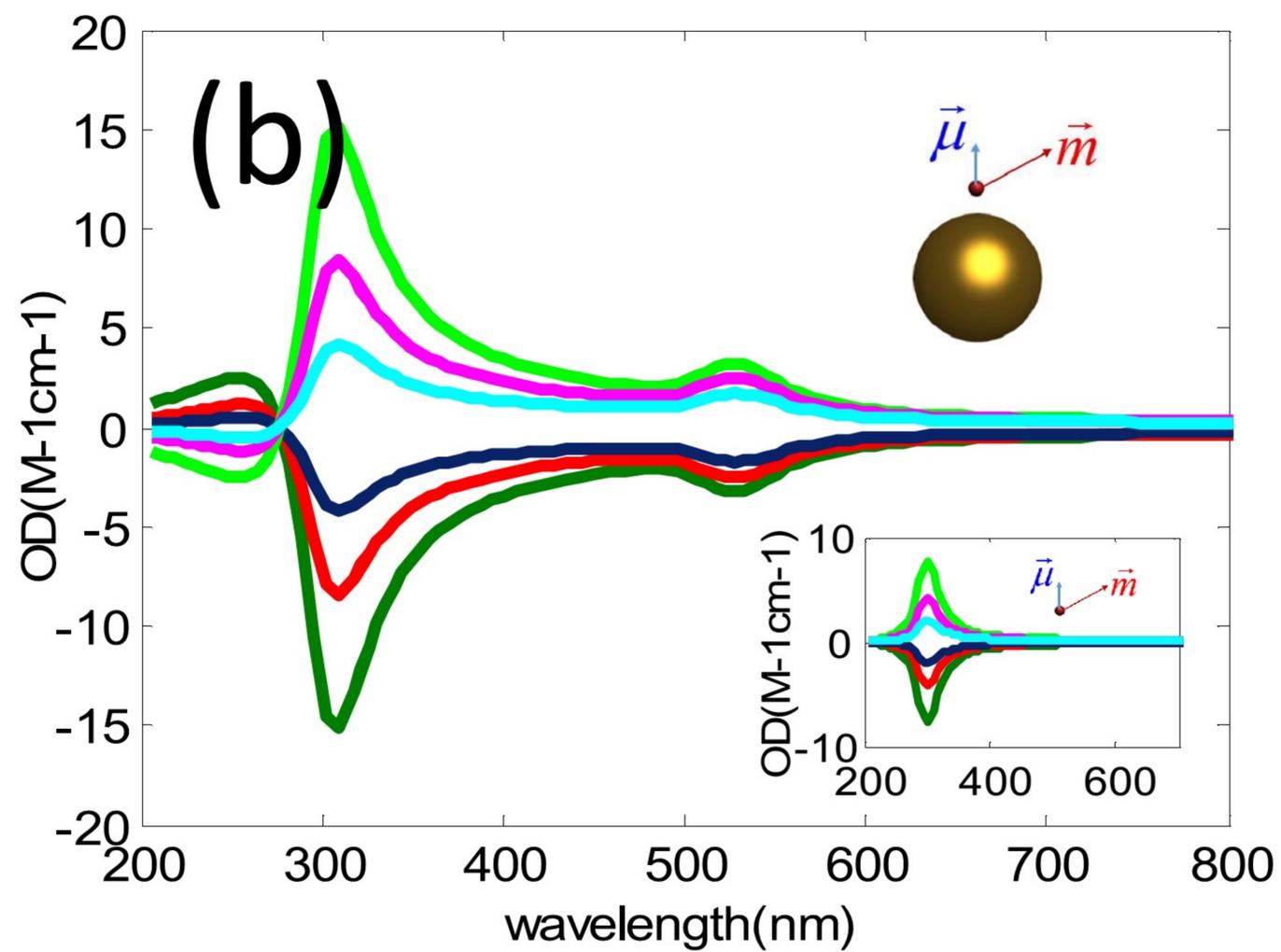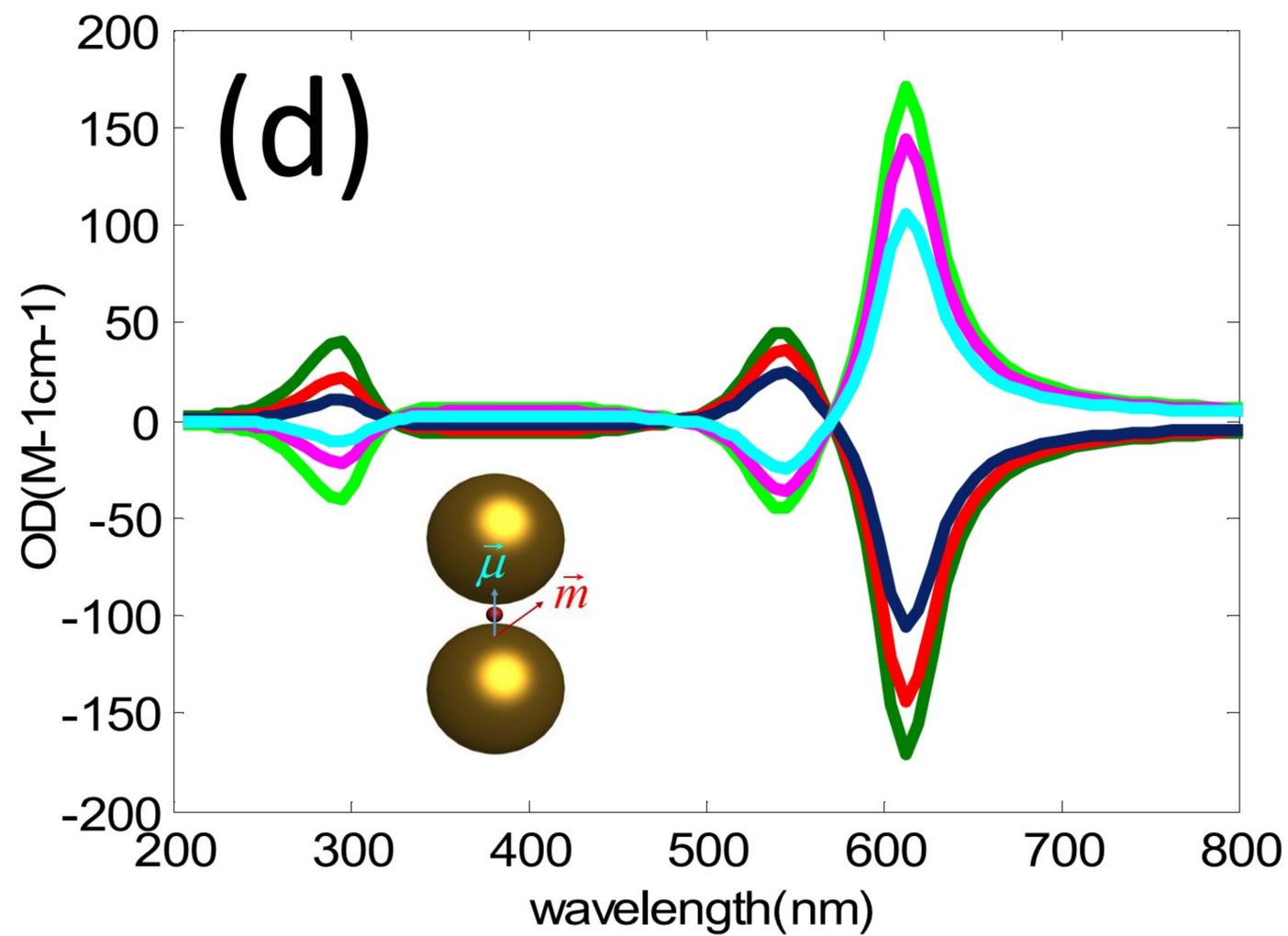